\documentclass[aps,superscriptaddress,nofootinbib]{revtex4-2}

\usepackage[T1]{fontenc}
\usepackage[utf8]{inputenc}
\usepackage{amsmath,amssymb,mathtools,bm}
\usepackage{graphicx}
\usepackage{hyperref}
\usepackage{slashed}

\usepackage{physics}

\usepackage{xcolor}

\newcommand{\dhd}{{\textstyle d}
	\lower.03ex\hbox{\kern-0.38em$^{\scriptstyle-}$}\kern-0.05em{}}
\newcommand{\dbar}{{\textstyle \delta}
	\lower.03ex\hbox{\kern-0.38em$^{\scriptstyle-}$}\kern-0.05em{}}
\newcommand{\half}{{1\over 2}}

\newcommand{\barq}{{\bar q}}

\newcommand{\baru}{{\bar u}}

\newcommand{\calf}{{\cal F}}

\newcommand{\calm}{{\cal M}}
 
\newcommand{\calo}{{\cal O}} 
 
\newcommand{\calq}{{\cal Q}}

\newcommand{\calu}{{\cal U}}

\newcommand{\barpsi}{{\bar \psi}}

\newcommand{\hatp}{{\hat p}}

\newcommand{\tildeD}{{\tilde D}}

\newcommand{\tildeQ}{{\tilde Q}}

\newcommand{\ketx}{\ket{x}}
\newcommand{\kety}{\ket{y}}

\newcommand{\brax}{\bra{x}}

\newcommand{\ketyp}{\ket{y_\perp}}

\newcommand{\braxp}{\bra{x_\perp}}

\newcommand \sslash [1] {\slash\hspace{-0.2cm}{#1}}
\newcommand{\ssp}{\sslash{p}}
\newcommand{\ssn}{\sslash{n}}
\newcommand{\ssk}{\sslash{k}}

\newcommand{\ssx}{\sslash{x}}
\newcommand{\ssy}{\sslash{y}}

\newcommand{\slashd}{\sslash{\partial}}

\newcommand \Slash [1] {\slash\hspace{-0.23cm}{#1}}

\newcommand{\Sp}{\Slash{P}}

\begin{document}
	
\title{From Sub-eikonal DIS to Quark Distributions and their High-Energy Evolution}
	
	\author{Giovanni Antonio Chirilli}
	\affiliation{Theoretical Physics Division, National Centre for Nuclear Research, Pasteura 7, Warsaw 02-093, Poland}
	
	\begin{abstract}
Relating the high-energy dipole description of deep-inelastic scattering to the standard light-ray operator formulation at finite Bjorken $x_B$ is essential for connecting the 
small-$x$ framework to the usual partonic description. I demonstrate that this connection already emerges at the first sub-eikonal order. 
At the differential level, the first 
sub-eikonal correction is governed by a quark TMD-like light-ray operator. In the inclusive limit, after complete phase-space integration, it reconstructs the standard nonlocal 
quark and helicity distributions at nonzero $x_B$. 
I then show independently that the same inclusive operator content follows from the high-energy 
limit of the leading-twist non-local operator product expansion, thereby establishing an explicit operator-level bridge between the shock-wave formalism 
and the non-local light-cone expansion.

I further discuss the high-energy evolution of the corresponding operators at $x_B=0$. Rewriting the evolution equations in terms of dipole-type operator 
combinations, I identify an operator basis whose bilocal building blocks vanish in the zero-dipole-size limit, 
making the small-dipole behavior and the leading-logarithmic structure 
manifest. In the double-logarithmic approximation the evolution equations admit the usual mixed longitudinal-transverse 
Bessel-type solution when the transverse phase space is treated independently. When the transverse phase space is instead constrained by longitudinal ordering, the second 
logarithm is converted into a logarithm of energy, and in the symmetric double-logarithmic regime one recovers the fixed-coupling Kirschner-Lipatov exponent with the full 
finite-$N_c$ color factor $C_F$.

	\end{abstract}
	
	\maketitle
	
\section{Introduction}

At high energy, deep-inelastic scattering (DIS) is naturally described in terms of Wilson lines~\cite{Balitsky:1995ub}
and the dipole picture~\cite{Nikolaev:1990ja, Nikolaev:1991et, Mueller:1993rr},
whereas at finite Bjorken $x_B$ it is formulated in terms
of nonlocal light-ray operators and parton distributions~\cite{Balitsky:1987bk,Balitsky:1990ck,Collins:2011zzd}.
Understanding how these two descriptions are connected is a basic question in QCD.
This issue becomes especially important in the small-Bjorken-$x_B$ regime, where perturbation theory is enhanced by large logarithms of the energy, or equivalently of $1/x_B$, whose resummation is governed by the Balitsky-Fadin-Kuraev-Lipatov (BFKL) evolution equation~\cite{Kuraev:1977fs,Balitsky:1978ic}.
The corresponding growth of gluon densities eventually drives the system toward the saturation regime, where multiple scattering and gluon recombination become indispensable~\cite{Gribov:1983ivg,Mueller:1985wy, McLerran:1993ka}.
In DIS, this region is naturally formulated in the dipole picture~\cite{Nikolaev:1990ja, Nikolaev:1991et, Mueller:1993rr}, in which the virtual photon fluctuates into a quark-antiquark pair that subsequently interacts with the target background through Wilson lines.
The phenomenological relevance of this regime was made clear by the DIS measurements at HERA, whose combined inclusive data still provide the standard benchmark for analyses of QCD dynamics at small $x$~\cite{H1:2015ubc}.
Beyond the strict high-energy limit, one expects power-suppressed corrections in the high-energy expansion to carry precisely the operator information that is absent in the dipole approximation.

This question is particularly timely in view of the Electron-Ion Collider, where one expects a broad kinematic region in which the separation between the asymptotic
small-$x_B$ regime and the conventional finite-$x_B$ partonic regime is no longer adequate~\cite{AbdulKhalek:2021gbh}.
From the theoretical point of view, one would like to understand how the familiar quark and helicity light-ray operators of the partonic description emerge from the 
Wilson-line framework when one moves beyond the eikonal approximation.
Sub-eikonal corrections provide the natural setting for this problem: they encode the first energy-suppressed interactions absent in the dipole limit and at the same time 
furnish the bridge between the high-energy expansion and the standard light-cone operator description.
A variety of approaches have been developed to address this problem, and the associated literature is by now quite broad~\cite{Balitsky:2015qba,Balitsky:2016dgz, 
Agostini:2019avp, Balitsky:2019ayf, Balitsky:2022vnb, Altinoluk:2020oyd,  Mukherjee:2023snp, Altinoluk:2025ang, Altinoluk:2025ivn, Mukherjee:2026cte}.

A closely related motivation comes from polarized scattering.
In the strict eikonal approximation the Wilson-line interaction is spin-blind, so genuinely helicity-sensitive observables cannot be described at leading power in the high-energy limit.
One therefore expects the first nontrivial quark and helicity operator content to appear precisely at sub-eikonal order.
The problem is then not only to identify the corresponding operators, but also to understand their evolution at high energy and their relation to the standard light-ray distributions at finite $x_B$.

The small-$x$ evolution of valence-quark quantum numbers in a dipole-Reggeon framework, including saturation effects, was studied in Ref.~\cite{Itakura:2003jp}.
The focus there was on valence-quark distributions and baryon-number transport at small $x$, whereas the present work is concerned with the operator-level relation 
between sub-eikonal DIS, the standard finite-$x_B$ quark and helicity light-ray distributions, and the corresponding high-energy evolution.

In this work we address this problem from two complementary directions.
First, we show that the first sub-eikonal quark correction to the dipole framework already reconstructs, in the inclusive limit, the standard quark and helicity light-ray 
distributions at nonzero Bjorken $x_B$.
At the differential level, the same correction is governed by a quark transverse-momentum-dependent light-ray operator.
We further show that the corresponding inclusive operator content is reproduced independently from the high-energy limit
of the non-local OPE.
In this way, the relation between the two operator formalisms becomes explicit:
the straight gauge-link structure of the non-local OPE reorganizes, in the high-energy limit, into the Wilson-line structures natural in the shock-wave description.
This provides an independent derivation of the same finite-$x_B$ quark operator content and establishes an explicit operator-level bridge between the high-energy expansion and the non-local light-cone formalism.

Second, we discuss the high-energy evolution of the corresponding $x_B=0$ operators.
In the approximation in which the fields are ordered according to the Sudakov component $k^+$, the relevant quark and helicity operators are $Q^f_1$ and $Q^f_5$.
We rewrite their evolution in terms of dipole-type operator combinations that vanish in the zero-dipole-size limit,
making the structure responsible for the leading logarithm of energy explicit.
This form is particularly useful for separating the genuinely dipole-like contributions from the local pieces and for clarifying the operator basis appropriate to the high-energy problem.

We then study the double-logarithmic approximation of the evolution equations of $Q^f_1$ and of the non-singlet $Q^{\rm NS}_5$.
A central point is that two different double-logarithmic regimes must be distinguished.
If the transverse phase space is treated as independent of the longitudinal evolution variable, one obtains the usual mixed longitudinal-transverse
resummation of powers of $(\alpha_s \ln(1/x_B)\ln Q_\perp^2)^n$, with the corresponding Bessel-type solution.
By contrast, if the transverse phase space is constrained by longitudinal ordering, the second logarithm becomes again a logarithm of energy.
In this regime the same ladder reproduces the fixed-coupling Kirschner-Lipatov exponent in the
symmetric double-logarithmic region.
The same reasoning applies to $Q^{\rm NS}_5$ in the non-singlet ladder approximation, while differences with the full polarized small-$x$ resummation
are expected beyond the ladder sector.

The purpose of the present paper is therefore twofold.
On the one hand, it makes explicit how the first sub-eikonal quark correction in the high-energy formalism already reconstructs the standard finite-$x_B$ quark and helicity light-ray operators.
On the other hand, it provides the corresponding high-energy evolution framework for the associated $x_B=0$ operators and clarifies the relation between the mixed double-logarithmic regime and the genuine double logarithm of energy.

The paper is organized as follows.
In Sec.~\ref{sec:DISsubeik1pointinSW} we derive the quark and helicity operator structures from the high-energy shock-wave formalism
and discuss the noncommutativity of the high-energy approximation and the full phase-space integration.
In Sec.~\ref{sec:nonOPE2heOPE} we recover the same inclusive operator content from the high-energy limit of the non-local OPE.
In Sec.~\ref{sec:evolutionQ1Q5} we discuss the high-energy evolution of the operators $Q_1$ and $Q_5$,
rewrite it in dipole form, and solve its double-logarithmic approximation,
including the case with the kinematic constraint.
 
\section{DIS at sub-eikonal level}
\label{sec:DISsubeik1pointinSW}

It is well known that the quark contribution to the structure functions is
energy suppressed at small \(x_B\).  A simple way to see this is to consider
the small-\(z\) behavior of the DGLAP splitting functions, where \(z\) is the
partonic momentum fraction.  At leading order \(P_{qq}^{(0)}(z)\) is finite
as \(z\to0\), whereas the gluonic kernels contain a \(1/z\) enhancement.
Thus,
\begin{eqnarray}
{P_{qq}^{(0)}(z)\over P_{gg}(z)} \xrightarrow[z\to 0]{} 0\,, \quad\quad {P_{qq}^{(0)}(z)\over P_{gq}(z)} \xrightarrow[z\to 0]{} 0\,.
\label{qq-smallx}
\end{eqnarray}
Since in this paper we are interested precisely in sub-eikonal, \textit{i.e.} energy-suppressed,
corrections, in this section we will derive the quark contribution to DIS at small $x_B$.

We will derive the corresponding quark distributions in two different ways. 
First, we adopt the high-energy Wilson-line formalism with the quark propagator in the shock wave.
In this case, unlike in the eikonal dipole approximation, the relevant quark propagator has one point inside the shock wave and the other outside. 
The second method, developed in the next section, is based on the non-local OPE in the collinear limit, 
followed by the high-energy boost that reproduces the result obtained in the first method.

\subsection{DIS at sub-eikonal level: Diagrams with 1-point in the shock-wave}

In this subsection we proceed in three steps. First, we calculate the transition amplitudes
with one point in the shock wave and project them onto longitudinal and transverse photon
polarizations. Second, from the differential cross section we isolate the quark-TMD
structure (for reviews see refs.~\cite{Angeles-Martinez:2015sea, Collins:2011zzd}). 
Third, after integrating over the final-state phase space, we discuss the inclusive
limit and show that the high-energy limit and the complete phase-space integration do not
commute. As a consequence, one is naturally led to distinguish between the naive collinear
quark pdf at $x_B=0$ and the quark distribution defined through light-ray operators.

\begin{figure}[h]
	\begin{center}
		\includegraphics[width=4 in]{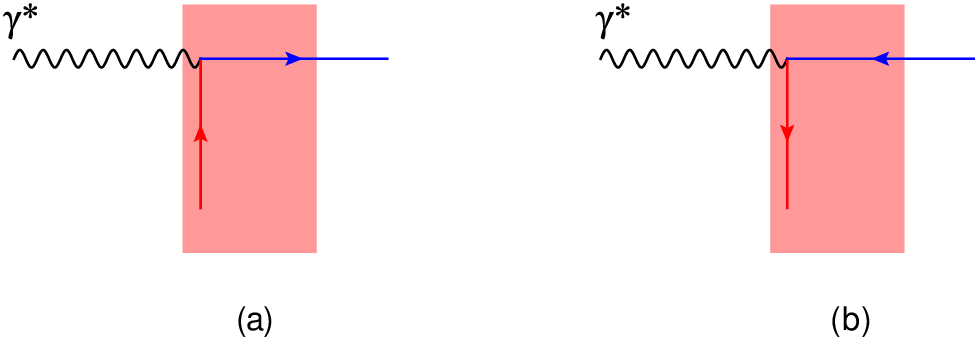}
		\caption{Diagrams contributing to the transition amplitude $\gamma^*(q)\to q(k)$ in the left panel, and $\gamma^*(q)\to \barq(k)$
			in the right panel. The blue fermionic lines are the quantum fields, while the red ones are the
			classical fields.}
		\label{Fig:DIS-q-1pointSW}
	\end{center}
	
\end{figure}

The minimal configuration in which the quark operator content of DIS enters beyond the strict dipole approximation 
is when one endpoint of the quark propagator lies inside the shock wave.  
For this reason, the diagrams in Fig.~\ref{Fig:DIS-q-1pointSW}, in which the virtual photon produces either a quark or an antiquark in the external field, provide the relevant starting point for the sub-eikonal analysis. 
We treat these contributions in the background-field method: first the amplitudes are computed in the shock-wave field, then the corresponding operator structures are isolated, and finally the resulting matrix elements are evaluated in the target state.

We start from the projection of the hadronic tensor onto the photon polarization vectors.
Restricting ourselves to the one-particle final states, it is given by the sum of the
contributions in which the virtual photon produces either a quark or an antiquark in the
background field. Thus, we have

\begin{eqnarray}
&&\hspace{-0.7cm}\varepsilon^\mu(q)\varepsilon^{*\nu}(q)W_{\mu\nu}
= {1\over 4\pi e^2}{1\over 2\pi\delta(0)}
\int \dhd^4k\dbar(k^2) 
\nonumber\\
&&\hspace{2.3cm}\times\Big(\left|\langle q(k)|\gamma^*(q) N(P,S)\rangle \right|^2
+ \left|\langle \barq(k)|\gamma^*(q) N(P,S)\rangle \right|^2\Big)\,.
\label{Wmunu}
\end{eqnarray}
where we divided by the infinite light-cone-volume $2\pi\delta(0)$ which comes from the delta-function square.

The hadronic tensor $W_{\mu\nu}$ is
\begin{eqnarray}
\hspace{-1cm}W_{\mu\nu}=\!\!&& \left(- g_{\mu\nu} + {q_\mu q_\nu\over q^2}\right)F_1(x_B, Q^2)
+ \left(P_\mu - q_\mu {q\cdot P\over q^2}\right) \left(P_\nu - q_\nu {q\cdot P\over q^2}\right){F_2(x_B, Q^2)\over P\cdot q}
\nonumber\\
\hspace{-1cm}&& + i\,\epsilon_{\mu\nu\lambda\sigma}\,q^\lambda S^\sigma {M\over P\cdot q}\,g_1(x_B,Q^2)
+  i\,\epsilon_{\mu\nu\lambda\sigma}q^\lambda\left(S^\sigma  - P^\sigma {q\cdot S\over q\cdot P}\right){M\over q\cdot P}\,g_2(x_B,Q^2)
\label{Wmunu-a}
\end{eqnarray}
and the virtual photon momentum is parameterized as
\begin{eqnarray}
q^\mu = q^+ n_1^\mu + q^- n_2^\mu + q_\perp^\mu \, .
\end{eqnarray}
Here, $n_1^\mu$ and $n_2^\mu$ are two light-like vectors such that
\begin{eqnarray}
n_1^2 = n_2^2 = 0,
\qquad
n_1 \cdot n_2 = 1 \, .
\end{eqnarray}
With these conventions, we have
\begin{eqnarray}
\slashed{n}_1 = \gamma^-,
\qquad
\slashed{n}_2 = \gamma^+ \, .
\end{eqnarray}
For a transverse vector we use the notation
\begin{eqnarray}
p_\perp^\mu = (0,p^1,p^2,0),
\end{eqnarray}
so that
\begin{eqnarray}
p_\perp^\mu p_{\perp\mu} = p^i p_i = -(p^1)^2-(p^2)^2 \, .
\end{eqnarray}
For the Euclidean scalar product in the transverse plane, we use the notation
\begin{eqnarray}
(p,q)_\perp = p^1 q^1 + p^2 q^2 \, .
\end{eqnarray}

Throughout this work, we use the $\hbar$-inspired notation
\begin{eqnarray}
	\dhd^n k \equiv \frac{d^n k}{(2\pi)^n}\,,
	\qquad
	\dbar^{(n)}(k)\equiv (2\pi)^n \delta^{(n)}(k)\,,
\end{eqnarray}
so that
\begin{eqnarray}
	\int \dhd^n k\, \dbar^{(n)}(k)=1\,.
\end{eqnarray}

To calculate the two diagrams shown in Fig.~\ref{Fig:DIS-q-1pointSW}, we need the quark
propagator in the shock-wave formalism with one point in the external field and the other
one outside~\cite{Chirilli:2018kkw,Chirilli:2021lif}. We first consider the transition
amplitude $\gamma^\ast(q)\to q(k)$ in the background of the gluon field generated by the
hadronic target. In the first approximation, we assume that the target background is made
only of gluon fields. Following the logic of the high-energy OPE, we first compute the
matrix element in the background field, then isolate the relevant operator structures, and
only at the end evaluate them in the target state. So, we have
\begin{eqnarray}
\langle q(k)|\gamma^*(q)N(P)\rangle \to \langle q(k)|\gamma^*(q)\rangle_A
\end{eqnarray}
We will assume that the hadronic target carries a large $P^-$ component and that the virtual photon carries a large $q^+$ component. In this high-energy limit,
\begin{eqnarray}
s = (P+q)^2 \simeq 2P^- q^+ \, .
\end{eqnarray}

Following the logic of the high-energy OPE that we described above, we first calculate $\langle q(k)|\gamma^*_T(q)\rangle_A$,
isolate the relevant operators, the quark operator, and finally evaluate them in the target state. 

We will make use here of the Schwinger representation. For example, the free scalar propagator can be written as
\begin{eqnarray}
\brax \frac{i}{p^2 + i\epsilon} \kety = i\!\int\!\dhd^4 k \,{e^{-ik\cdot(x-y)}\over k^2 + i\epsilon}\,,
\label{schwrep}
\end{eqnarray}
with $\langle k\ketx = e^{ix\cdot k}$.

The matrix element for diagram in Fig. \ref{Fig:DIS-q-1pointSW}a is given by
\begin{eqnarray}
\hspace{-1.3cm} \langle q(k)|\gamma^*(q)\rangle_{\small{\rm Fig.}\ref{Fig:DIS-q-1pointSW}a}
=\!\!\!&& iee_f\left(\int d^4x\,\varepsilon_\mu(q)\,e^{-iq\cdot x}\right)
\nonumber\\
&& \times\!\!\int d^4y\, e^{ik\cdot y} \baru(k) \theta(k^+) i\slashd_y
\langle {\rm T}\{\psi_q(y)\barpsi_q(x)\gamma^\mu\psi_c(x^+,x_\perp)\}\rangle_A
\label{MEq-1pointSW-1}
\end{eqnarray}
where we indicated with the subscript $q$ the quantum fields and with $c$ the classical ones.
As usual, the classical field, subject to the high longitudinal boost, does not depend on the $x^-$ component.

\begin{figure}[t]
	\begin{center}
		\includegraphics[width=4.5in]{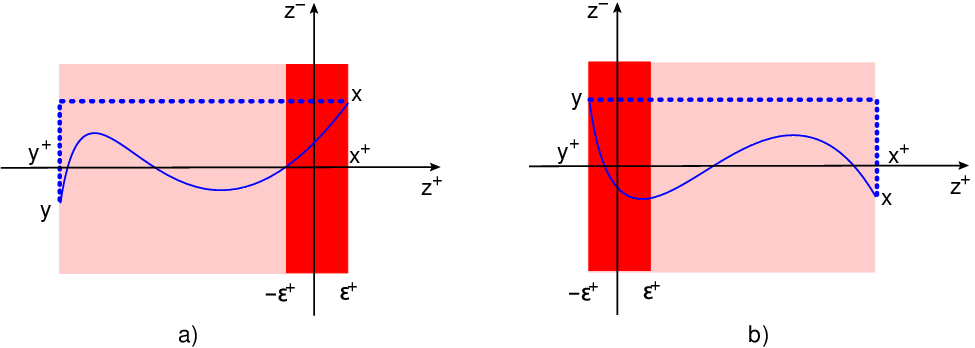}
		\caption{The quark starts its propagation in the shock-wave and ends it outside.
			In this case, the pure gauge is only on one side of the shock-wave. }
		\label{Fig:leftswprpagation}
	\end{center}
\end{figure}

Here, the point $x$ is in the background shock-wave field while the point y is outside. The quark propagator with one point in the shock-wave and the 
other one outside was calculated in ref.~\cite{Chirilli:2018kkw}
\begin{eqnarray}
&&\hspace{-0.5cm}
\brax \frac{i}{ \Sp+i\epsilon} \kety_{\small{\rm Fig.}\ref{Fig:leftswprpagation}a}
\nonumber\\
=\!\!\!&&\left[\int_0^{+\infty}\!\!{\dhd p^+\over 4(p^+)^2}\theta(x^+-y^+) - 
\int_{-\infty}^0\!\!{\dhd p^+\over 4(p^+)^2}\theta(y^+-x^+) \right]\! e^{-ip^+(x^--y^-)}
\nonumber\\
&&\hspace{-0.5cm}\times\braxp
\Big([x^+,y^+]\ssp - g\!\!\int_{y^+}^{x^+}\!\!\!d\omega^+\,\gamma^i\,[x^+,\omega^+]{F_i}^{\;-}
[\omega^+,y^+] \Big)
\ssn_2\ssp\,e^{i{\hatp^2_\perp\over 2p^+}(y^+-x^+)}\ketyp + O(\lambda^{-1})
\nonumber\\
=\!\!\!&&{1\over 2}\int\dhd^4k\Bigg[\theta(x^+-y^+)\Big([x^+,-\infty n_1]_x\ssk\ssn_2 - g\int_{-\infty}^{x^+}dz^+\gamma^i\ssn_2
[x^+,z^+]_x{F_i}^{\;-}(z^+,x_\perp)[z^+,-\infty n_1]_x\Big)
\nonumber\\
&&+\theta(y^+-x^+)\Big([x^+,\infty n_1]_x\ssk\ssn_2 + g\int^{+\infty}_{x^+}dz^+\gamma^i\ssn_2[x^+,z^+]_x{F_i}^{\;-}(z^+,x_\perp)[z^+,\infty n_1]_x\Big)
\Bigg]
\nonumber\\
&&\times\!{i\ssk\over k^+(k^2+i\epsilon)}\,e^{-ik\cdot(x-y)}
\label{xinSWfree-right}
\end{eqnarray}
The second equal sign is taken in the shock-wave limit where the limit of integration can be extended to infinity.
In the case of $y$ in the shock-wave, the propagator is
\begin{eqnarray}
&&\hspace{-0.5cm}\brax \frac{i}{\Sp+i\epsilon}\kety_{\small{\rm Fig.}\ref{Fig:leftswprpagation}b}
\nonumber\\
=\!\!\!&&\left[\int_0^{+\infty}\!\!{\dhd p^+\over 4(p^+)^2}\theta(x^+-y^+) - 
\int_{-\infty}^0\!\!{\dhd p^+\over 4(p^+)^2}\theta(y^+-x^+) \right]\! e^{-i p^+(x^--y^-)}
\nonumber\\
&&\hspace{0.5cm}\times\braxp\ssp\ssn_2\,e^{-i{\hatp^2_\perp\over 2p^+}(x^+-y^+)}
\Big(\ssp[x^+,y^+]
+ g\!\!\int_{y^+}^{x^+}\!\!\!d\omega^+\,\gamma^i\,[x^+,\omega^+]{F_i}^{\;-}
[\omega^+,y^+] \Big)\ketyp + O(\lambda^{-1})
\nonumber\\
=\!\!\!&&{1\over 2}\int\dhd^4k{i\ssk\over k^+(k^2+i\epsilon)}\,e^{-ik\cdot(x-y)}
\nonumber\\
&&\times\Bigg[\theta(x^+-y^+)\Big(\ssn_2\ssk[\infty n_1,y^+]_y + g\int^{+\infty}_{y^+}dz^+\ssn_2\gamma^i
[\infty n_1,z^+]_y{F_i}^{\;-}[z^+,y^+]_y\Big)
\nonumber\\
&&+\theta(y^+-x^+)\Big(\ssn_2\ssk[-\infty n_1,y^+]_y - g\int_{-\infty}^{y^+}dz^+\ssn_2\gamma^i[-\infty n_1,z^+]_y{F_i}^{\;-}[z^+,y^+]_y\Big)
\Bigg]
\label{xinSWfree-left}
\end{eqnarray}

Using propagator (\ref{xinSWfree-left}), we apply the LSZ reduction formula differentiating the theta-function and the exponential as explained 
in ref.~\cite{Chirilli:2026vij}, and from eq. (\ref{MEq-1pointSW-1}) we have
\begin{eqnarray}
&&  \langle q(k)|\gamma^*(q)\rangle_{\small{\rm Fig.}\ref{Fig:DIS-q-1pointSW}a}
\nonumber\\
=\!\!\!&& - {ee_f\over 2}\left(\int d^4x\,e^{-iq\cdot x}\varepsilon_\mu(q)\right) \int d^4y \, e^{ik\cdot y}\baru(k)\theta(k^+)
\int \dhd^4 k_1\,{e^{-ik_1\cdot(y-x)}\over k^+_1(k_1^2+i\epsilon)}
\nonumber\\
&&\times\Bigg[
\Big(i\ssn_2\delta(x^+-y^+) + \ssk_1\theta(y^+-x^+)\Big)\ssk_1\ssn_2
\nonumber\\
&&~~~\times\Big(\ssk_1[+\infty n_1,x^+]_x + g\int^{+\infty}_{x^+}dw^+\gamma^i[+\infty n_1,w^+]_x{F_i}^{\;-}[w^+,x^+]_x\Big)
\nonumber\\
&&\hspace{-0.5cm} - \Big(i\ssn_2\delta(x^+-y^+) - \ssk_1\theta(x^+-y^+)\Big)\ssk_1\ssn_2
\nonumber\\
&&\times\Big(\ssk_1[-\infty n_1,x^+]_x - g\int_{-\infty}^{x^+}dw^+\gamma^i[-\infty n_1,w^+]_x{F_i}^{\;-}[w^+,x^+]_x\Big)\Bigg]\gamma^\mu\psi(x^+,x_\perp)
\label{MEq-1pointSW-2}
\end{eqnarray}
The integration over $y^-$ gives $k^+_1= k^+$, the one over $y_\perp$ sets $k_{1\perp}=k_\perp$, 
and finally integrating over $y^+$ we arrive at
\begin{eqnarray}
&&\hspace{-1cm}\langle q(k)|\gamma^*(q)\rangle_{\small{\rm Fig.}\ref{Fig:DIS-q-1pointSW}a}
\nonumber\\
&&\hspace{-2cm}= i  ee_f\left(\int d^4x\,e^{-iq\cdot x}\varepsilon_\mu(q)\right) {i\over 2}\int\dhd k^-_1
{e^{-i(k,x)+ik^+ x^-}\over k^+(2k^+k^-_1-k^2_\perp+i\epsilon)}
\baru(k)\theta(k^+)\,i\,e^{ik^-_1x^+}
\nonumber\\
&&\hspace{-2cm}
\times\!\left(\ssn_2 - {\ssk_1\over k^-_1 - k^--i\epsilon}\right)(k^+\ssn_1+\ssk_\perp)\ssn_2
\Big((k^+\ssn_1+\ssk_\perp)[+\infty n_1,x^+]_x 
\nonumber\\
&&\hspace{-2cm} ~~~~ + g\int^{+\infty}_{x^+}dw^+\gamma^i[+\infty n_1,w^+]_x{F_i}^{\;-}[w^+,x^+]_x\Big)\gamma^\mu\psi(x^+,x_\perp)
\label{MEq-1pointSW-3}
\end{eqnarray}
Note that to arrive at result (\ref{MEq-1pointSW-3}) we have used the fact that 
the integration over $x^-$, will set $q^+=k^+$, so, the second term in (\ref{MEq-1pointSW-2}) does not contribute
because $q^+>0,$ and the calculation of the residue in $k^-_1$ is zero. 

The final step is to calculate the residue integrating over $k_1^-$, and obtain
\begin{eqnarray}
	&& \langle q(k)|\gamma^*(q)\rangle_{\small{\rm Fig.}\ref{Fig:DIS-q-1pointSW}a}
	\nonumber\\
	=\!\!\!&& -{1\over 2}\dbar(q^+ - k^+)ee_f{\baru(k)\theta(k^+)\over k^+}
	\int d^2x dx^+ e^{-i(q^- - k^-)x^+ + i(q-k,x)_\perp}\ssn_2
	\Big((k^+\ssn_1+\ssk_\perp)[+\infty n_1, x^+]_x 
	\nonumber\\
	&&
	+ g \int_{x^+}^{+\infty}\!\! dw^+\gamma^i[+\infty n_1,w^+]_x {F_i}^{\;-}[w^+,x^+]_x\Big)\sslash{\varepsilon}(q)\psi(x^+,x_\perp)
	\label{MEq-1pointSWa}
\end{eqnarray}
and for diagram in Fig. \ref{Fig:DIS-q-1pointSW}b, we have (see appendix \ref{sec:FigDIS-q-1pointSWb})
\begin{eqnarray}
	&& \langle \bar{q}(k)|\gamma^*(q)\rangle_{\small{\rm Fig.}\ref{Fig:DIS-q-1pointSW}b}
	\nonumber\\
	=\!\!\!&& - {1\over 2}\dbar(q^+ - k^+)ee_f\,{\theta(k^+)\over k^+}\int \!\! d^2xdx^+ e^{-i(q^- - k^-)x^+ + i(q-k,x)_\perp}
	\barpsi(x^+,x_\perp)\sslash{\varepsilon}(q)
	\nonumber\\
	&&\times\Big([x^+,\infty n_1]_x(k^+ \ssn_1+\ssk_\perp) - g\!\int_{x^+}^{+\infty} \!dw^+\gamma^i[x^+,w^+]_x{F_i}^{\;-}[w^+,\infty n_1]_x\Big)\ssn_2v(k)
	\label{MEq-1pointSWb}
\end{eqnarray}
Before squaring the matrix element (\ref{MEq-1pointSWa}), and (\ref{MEq-1pointSWb}), 
let us consider their projections onto longitudinal and transverse photon polarizations.

\subsubsection{Longitudinal photon polarization}

Let us start with the longitudinal polarization
$\varepsilon^L_\mu = {q^+\over Q}n_{1\mu} + {Q\over 2q^+}n_{2\mu}$ with  $Q= \sqrt{-q^2}$. 
In eq. (\ref{MEq-1pointSWa}), we have the following two Dirac matrix elements
\begin{eqnarray}
	&&\ssn_2(k^+\ssn_1+\ssk_\perp)\sslash{\varepsilon}^L(q)\psi(x^+,x_\perp)
	\nonumber\\
	=\!\!\!&&Q\ssn_2\psi(x^+,x_\perp) - {s\over 2q^+Q}k_i(\gamma^i-i\epsilon^{ij}\gamma_j\gamma^5)\psi(x^+,x_\perp)
	\nonumber\\
	=\!\!\!&& - {q^+\over Q}k_i(\gamma^i-i\epsilon^{ij}\gamma_j\gamma^5)\psi(x^+,x_\perp) + O(\lambda^{-2})
	\label{ME-longy1}
\end{eqnarray}
and
\begin{eqnarray}
	\ssn_2\gamma^i\sslash{\varepsilon}^L(q)\psi(x^+,x_\perp) = - {q^+\over Q}(\gamma^i-i\epsilon^{ij}\gamma_j\gamma^5)\psi(x^+,x_\perp)
	\label{ME-longy2}
\end{eqnarray}
Using result (\ref{ME-longy1}) and (\ref{ME-longy2}), the longitudinal contribution to the amplitude (\ref{MEq-1pointSWa}) is
\begin{eqnarray}
	&&\langle q(k)|\gamma^*_L(q)\rangle_{\small{\rm Fig.}\ref{Fig:DIS-q-1pointSW}a}
	\nonumber\\
	=\!\!\!&& {ee_f\over 2 Q}\int\!d^2 x_\perp dx^+\, e^{i(q-k,x)_\perp - i(q^--k^-)x^+}\dbar(q^+-k^+)  \baru(k)\theta(k^+)
	\nonumber\\
	&&\times(\gamma^i-i\epsilon^{ij}\gamma_j\gamma^5)\Big(k_i[\infty n_1, x^+]_x + \big(-i\mathfrak{D}_i[\infty n_1, x^+]_x\big)\Big)\psi(x^+,x_\perp)
	+ O(\lambda^{-2})
	\nonumber\\
	=\!\!\!&& {ee_f\over 2 Q}\int\!d^2 x_\perp dx^+\, e^{i(q-k,x)_\perp - i(q^- -k^-)x^+}\dbar(q^+-k^+)  \baru(k)\theta(k^+)
	\nonumber\\
	&&\times(\gamma^i-i\epsilon^{ij}\gamma_j\gamma^5)\Big(-q_i[\infty n_1, x^+]_x\psi(x^+,x_\perp) + [\infty n_1, x^+]_xD_i^x\psi(x^+,x_\perp)\Big)
	\nonumber\\
	&&+ O(\lambda^{-2})
	\label{MEq-1pointSW-1a}
\end{eqnarray}
where we have used the covariant derivative on the Wilson line (see appendix of ref.~\cite{Chirilli:2018kkw}).

Now observe that the square of the longitudinal matrix element, eq.~\eqref{MEq-1pointSW-1a}, is sub-sub-eikonal compared with contributions of the type $\bar\psi \slashed{n}_1 \psi$. For this reason, we will disregard it.

\subsubsection{Transverse photon polarization}

Here we consider the transverse polarization using $\varepsilon^\lambda_k = -{1\over \sqrt{2}}(\lambda,i)$ with $\lambda=\pm 1$.
We are interested in sub-eikonal corrections so we will neglect all sub-leading contribution.

Applying the approximations (\ref{spinorboost}), the two terms in the matrix element (\ref{MEq-1pointSWa}) become
\begin{eqnarray}
\ssn_2(k^+\ssn_1+\ssk_\perp)\gamma^i\psi = && \Big[k^+(\gamma^i-i\epsilon^{ij}\gamma_j\gamma^5)
+ \ssn_2(g^{ij}-i\epsilon^{ji}\gamma^5)k_j\Big]\psi
\nonumber\\
= && k^+(\gamma^i-i\epsilon^{ij}\gamma_j\gamma^5)\psi + O(\lambda^{-2})
\label{MEq-approxyTermA}
\end{eqnarray}
and
\begin{eqnarray}
{F_i}^{\;-}[w^+,x^+]_x\ssn_2\gamma^i\gamma^j\varepsilon_j^\lambda\psi = 
{F_i}^{\;-}[w^+,x^+]_x\varepsilon_j^\lambda\ssn_2(g^{ij} - i\epsilon^{ij}\gamma^5)\psi\sim\lambda^{-2}
\label{MEq-approxyTermB}
\end{eqnarray}
So, using (\ref{MEq-approxyTermA}) and (\ref{MEq-approxyTermB}), the transverse component of the matrix element (\ref{MEq-1pointSWa}), is
\begin{eqnarray}
&&\langle q(k)|\gamma^*_T(q)\rangle_{\small{\rm Fig.}\ref{Fig:DIS-q-1pointSW}a}
\nonumber\\
=\!\!\!&& -{ee_f\over 2}\dbar(q^+-k^+)\baru(k)\theta(k^+) \!\!\int d^2 x_\perp\, e^{i(q-k,x)_\perp}
\!\!\int dx^+ e^{i(k^--q^-)x^+}
\nonumber\\
&&\times(\sslash{\varepsilon}^\lambda_\perp - i \varepsilon^\lambda_i\epsilon^{ij}\gamma_j\gamma^5)
[+\infty n_1, x^+]_x\psi(x^+,x_\perp) + O(\lambda^{-2})
\label{MEq-1pointSW-T}
\end{eqnarray}
We can now square the transverse component of the proton-to-quark transition amplitude and integrate over the phase space.
Thus, using (\ref{MEq-1pointSW-T}), we have
\begin{eqnarray}
&&{1\over 2\pi\delta(0)}\int \dhd^4k\dbar(k^2)\Big|\langle q(k)
|\gamma^*_T(q)\rangle_{\small{\rm Fig.}\ref{Fig:DIS-q-1pointSW}a}\Big|^2
\nonumber\\
&&\hspace{-0.4cm} = \sum_{\sigma,f}\half\sum_{\lambda=\pm 1}
\Bigg|-{ee_f\over 2}\baru(k,\sigma)\theta(k^+)\!\!\int d^2 x_\perp\, e^{i(q-k,x)_\perp}
\!\!\int dx^+ e^{-i(q^--k^-)x^+}
\nonumber\\
&&\times(\sslash{\varepsilon}^\lambda_\perp - i \varepsilon^\lambda_i\epsilon^{ij}\gamma_j\gamma^5)
[+\infty n_1, x^+]_x\psi(x^+,x_\perp)\Bigg|^2\dhd^4k\dbar(k^2)\dbar(q^+-k^+) +O(\lambda^{-2})
\nonumber\\
&&\hspace{-0.4cm}
= {1\over 2}\sum_f e^2e^2_f\int d^2x d^2y\dhd^2k e^{i(q-k,x-y)_\perp}\int dx^+ dy^+ 
\, e^{-i\left(q^- - {k^2_\perp\over 2q^+}\right)(x^+-y^+)}
\nonumber\\
&&\hspace{-0.2cm}
~~\times\!\barpsi(y^+,y_\perp)[y^+,+\infty n_1]_y\ssn_1[+\infty n_1, x^+]_x\psi(x^+,x_\perp) +O(\lambda^{-2})
\label{squMEq-1pointSW-Ta}
\end{eqnarray}
where we indicated with $2\pi\delta(0)$ the infinite light-cone-volume which comes from the delta-function square.
In eq. (\ref{squMEq-1pointSW-Ta}), we can use $q^- = -{Q^2\over2q^+}=-x_BP^-$ at $q_\perp=0$, and 
\begin{eqnarray}
	(\gamma^i+i\epsilon^{il}\gamma_l\gamma^5)\ssn_1(\gamma^i-i\epsilon^{ij}\gamma_j\gamma^5) = 8\ssn_1
\end{eqnarray}
with $\sum_{\lambda=\pm 1}\varepsilon^\lambda_i\varepsilon^{\lambda *}_j = \delta_{ij}$.

From the result above, we will first derive the differential cross section for the
$\gamma^*\to q$ process and thus obtain the quark TMD. We will then consider the
inclusive cross section. In doing so, we will show that the small-$x_B$ limit and the
complete integration over the phase space do not commute. These two operations lead to
two different quark distributions, namely the naive collinear quark pdf at $x_B=0$ and the
quark distribution defined through light-ray operators.

\subsubsection{Quark-TMD cross-section}

Let us now consider the forward matrix element of the result above. In this way, the
differential cross section can be written directly in terms of a nonlocal quark operator in the
target state, which is the natural object underlying the quark TMD at small $x_B$. 
We define the reduced matrix element as
\begin{eqnarray}
\langle\langle\calo\rangle\rangle = \lim_{\epsilon^+\to 0}{\bra{N(P)}\calo\ket{N(P+\epsilon^+ n_1)}\over \dbar(\epsilon^+)\dbar^{(2)}(0)}
\label{reducedME}
\end{eqnarray}
Thus, the forward matrix element of result (\ref{squMEq-1pointSW-Ta}), at $q_\perp=0$, becomes
\begin{eqnarray}
&&{1\over 2\pi\delta(0)}\int \dhd^4k\dbar(k^2)\Big|\langle q(k)
|\gamma^*_T(q)\rangle_{\small{\rm Fig.}\ref{Fig:DIS-q-1pointSW}a}\Big|^2
\nonumber\\
=\!\!\!&&\lim_{\epsilon^+\to 0}{1\over 2} e^2e^2_f\,\int d\Delta^+ d^2r {d^2k\over (2\pi)^2}
e^{-i(k,r)_\perp}e^{i\left(x_B+{k^2_\perp\over s}\right)P^-\Delta^+}
\nonumber\\
&&\times
\bra{N(P)}\barpsi(0^+,0_\perp)\,[0^+,+\infty n_1]_{0_\perp}\,\ssn_1\,[+\infty n_1,\Delta^+]_{r}\,\psi(\Delta^+,r_\perp)\ket{N(P+\epsilon^+ n_1)}\nonumber\\
&&\equiv \int d^2k \langle\langle\calm^{LO}_{\rm qTMD}\rangle\rangle
\label{reduceTMDsqared}
\end{eqnarray}
with $\Delta^{\!+}=x^+-y^+$, $x_B={Q^2\over s}$.
In eq. \eqref{reduceTMDsqared}, we obtain the quark TMD at $x_B\ne 0$ (cf.~\cite{Bacchetta:2006tn})
\begin{eqnarray}
\hspace{-0.6cm}q_{1,f}(x_B,k_\perp) \equiv\!\!\!&& {1\over 4\pi}\int d\Delta^{\!+} d^2r \,
e^{-i(k, r)_\perp}e^{i\left(x_B+{k^2_\perp\over s}\right)P^-\Delta^+}
\nonumber\\
&&\times
\bra{N(P)}\barpsi_f(0^+,0_\perp)\,[0^+, +\infty n_1]_0\,\gamma^-\,[+\infty n_1,\Delta^{\!+}]_{r}\,\psi_f(\Delta^{\!+},r_\perp)\ket{N(P)}\,.
\label{qTMD}
\end{eqnarray}
Keeping the exact on-shell phase before taking the strict high-energy limit, one finds the kinematic factor
\begin{eqnarray}
\exp\!\left(i\,\frac{k_\perp^2}{s}\,P^- \Delta^+\right).
\end{eqnarray}
This factor drops out if the leading-power high-energy approximation is made before the TMD-like operator is identified.

Thus, the differential cross-section for quark-TMD is obtained from the reduced matrix element as
\begin{eqnarray}
{d^2\sigma^{\gamma^*_T\to q}\over d^2k} = 
{1\over 4P\cdot q}\langle\langle\calm^{LO}_{\rm qTMD}\rangle\rangle 
\label{cross-reduce}
\end{eqnarray}
So, the differential cross-section in terms of the quark-TMD $q_1(x_B,k_\perp)$, using (\ref{squMEq-1pointSW-Ta}) and (\ref{cross-reduce}), is
\begin{eqnarray}
&&{d^2\sigma^{\gamma^*_T\to q}\over d^2k} 
={\alpha_{\rm em}\over s}\,e^2_f\,q_{1,f}(x_B,k_\perp)
\end{eqnarray}
The whole nontrivial operator content is thus
contained in $q_{1,f}(x_B,k_\perp)$, which may be interpreted as the quark TMD-like
light-ray distribution generated by the first correction beyond the strict eikonal dipole
approximation. The semi-infinite Wilson lines encode the eikonal interaction with the
target background, while the Fourier transform in $\Delta^+$ and $r_\perp$ fixes the
longitudinal momentum fraction and transverse momentum of the final-state quark.
\subsubsection{Naive collinear quark pdf}

Before performing the full phase-space integration, it is instructive to observe what would happen if, in \eqref{squMEq-1pointSW-Ta}, one takes the high-energy limit too early, 
and sets
\begin{eqnarray}
e^{-i\Big(q^- - {k^2_\perp\over 2q^+}(x^+-y^+)\Big)} \to 1\,.
\label{HE-expoapproxy}
\end{eqnarray} 
Approximation \eqref{HE-expoapproxy} leads to the naive collinear quark pdf at $x_B=0$.
Indeed, the integration over the phase-space is a simple delta function which sets the two gauge links to the same transverse coordinate
and we recover the quark pdf. Indeed, from eq.~\eqref{squMEq-1pointSW-Ta}, we obtain
\begin{eqnarray}
&&\hspace{-1.4cm}{1\over 2\pi\delta(0)}\int \dhd^4k\dbar(k^2)\theta(k^+)
\Big|\langle q(k)|\gamma^*_T(q)\rangle_{\small{\rm Fig.}\ref{Fig:DIS-q-1pointSW}a}\Big|^2
\nonumber\\
&&\hspace{-1.8cm} 
= \half\sum_f e^2e^2_f\int d^2x dx^+dy^+
\,\langle\barpsi(y^+,x_\perp)\ssn_1[y^+,x^+]_x\psi(x^+,x_\perp)\rangle +O(\lambda^{-2})
\label{squMEq-1pointSW-Tb}
\end{eqnarray}
from which, using the translation invariance of the forward matrix element, we have the collinear quark pdf at $x_B=0$
\begin{eqnarray}
&&q_f(x_B=0) = {1\over 4\pi}\int dx^+\bra{N(P)}\barpsi_f(0^+,0_\perp)\gamma^-[0^+,x^+]\psi_f(x^+,0_\perp)\ket{N(P)}
\label{coll-qf}
\end{eqnarray}
The distribution (\ref{coll-qf}) is the usual collinear distribution at $x_B=0$. 
Similar conclusion was reached in ref.~\cite{Altinoluk:2025ang} where diagrams in Fig. \ref{Fig:DIS-q-1pointSW} were first considered.
In the next subsection, we are going to show that to obtain the quark distribution with $x_B$ dependence one has to complete 
the phase-space integration before taking the small-x limit.

\subsubsection{Quark pdf operator}

The proper inclusive small-$x_B$ limit is obtained only after the phase-space integration is
completed. In other words, the small-$x_B$ limit and the full phase-space integration do not
commute. If the small-$x_B$ approximation is performed too early, one recovers only the
naive parton-model result discussed above.

Therefore, we can complete the phase-space integration without making any small-$x_B$
approximation in the integrand. Integrating also over the transverse momentum $k_\perp$
in eq.~\eqref{squMEq-1pointSW-Ta}, we obtain
\begin{eqnarray}
&&{1\over 2\pi\delta(0)}\int \dhd^4k\dbar(k^2)\theta(k^+)
\Big|\langle q(k)|\gamma^*_T(q)\rangle_{\small{\rm Fig.}\ref{Fig:DIS-q-1pointSW}a}\Big|^2
\nonumber\\
=\!\!\!&& \half\sum_f e^2e^2_f\int d^2x d^2y \, e^{i(q,\Delta)_\perp}
\int dx^+dy^+\, {is\over 4\pi P^-\Delta^+}\,e^{-iq^-\Delta^+ - i{\Delta^2_\perp s \over 4P^-\Delta^+}}
\nonumber\\
&&\times\barpsi(y^+,y_\perp)[y^+,+\infty n_1]_y\ssn_1[+\infty n_1, x^+]_x\psi(x^+,x_\perp) 
+ O(\lambda^{-2})
\label{Int-squMEq-1pointSW-Ta}
\end{eqnarray}
where $\Delta=x-y$.

We are now ready to take the high-energy limit of the transverse kernel while keeping the longitudinal phase $e^{-iq^-\Delta^+}=e^{ix_B P^-\Delta^+}$
intact. In this sense, the following step is high-energy limit at fixed $x_B$, rather than the strict small-$x_B$ limit.
The transverse propagation is controlled by the kernel
\begin{eqnarray}
	K_s(\Delta_\perp,\Delta^+)
	\equiv{is\over 4\pi P^-\Delta^+ }
	\exp\!\left(
	-i{\Delta_\perp^2 s\over 4P^-\Delta^+}\right).
	\label{Ks-def}
\end{eqnarray}
For fixed $\Delta^+$, its Fourier transform is
\begin{eqnarray}
\int d^2\Delta_\perp\,
e^{i\ell_\perp\cdot \Delta_\perp}
K_s(\Delta_\perp,\Delta^+)
=\exp\!\left(i\,{\ell_\perp^2 P^-\Delta^+\over s}\right),
\end{eqnarray}
which tends to unity in the high-energy limit $s\to\infty$ at fixed $\Delta^+$.

Therefore, in the distributional sense under the $d^2\Delta_\perp$ integral, one has
\begin{eqnarray}
{is\over 4\pi P^-\Delta^+ }
\exp\!\left(
-i{\Delta_\perp^2 s\over 4P^-\Delta^+}\right)
\stackrel{s\to\infty}{\longrightarrow}
\delta^{(2)}(\Delta_\perp).
\label{delta}
\end{eqnarray}
The limit in eq.~\eqref{delta} shows that the first sub-eikonal correction already
reconstructs, in the inclusive case, the standard finite-$x_B$ nonlocal light-ray quark
operator. Indeed, eq.~\eqref{delta} localizes the operator in the transverse plane, while
the longitudinal phase $e^{-iq^-\Delta^+}=e^{ix_B P^-\Delta^+}$ remains exact and
preserves the Fourier structure conjugate to Bjorken $x_B$. Therefore, the last step is
taken at high-energy with fixed $x_B$.

Hence, the operator becomes local in the transverse plane, but remains nonlocal in the longitudinal direction, as appropriate for the standard light-ray quark operators at $x_B\neq 0$.
Using eq.~\eqref{delta} in eq.~\eqref{Int-squMEq-1pointSW-Ta}, and combining the two Wilson lines at coincident transverse position, we obtain
\begin{eqnarray}
&&{1\over 2\pi\delta(0)}\int \dhd^4k\dbar(k^2)\theta(k^+)
\Big|\langle q(k)|\gamma^*_T(q)\rangle_{\small{\rm Fig.}\ref{Fig:DIS-q-1pointSW}a}\Big|^2
\nonumber\\
=\!\!\!&&\half\sum_f e^2 e_f^2\int d^2x\,dx^+ dy^+\,e^{ix_B P^-\Delta^+}
\bar\psi_f(y^+,x_\perp)\,[y^+,x^+]_x\,\ssn_1\,\psi_f(x^+,x_\perp)
+ O(\lambda^{-2}),
\label{quarkSubEik-1pointinSWa}
\end{eqnarray}
with $\Delta^+=x^+-y^+$. Notice that, the results \eqref{quarkSubEik-1pointinSWa}, can be obtained directly from
eq.~\eqref{Int-squMEq-1pointSW-Ta} by Taylor expanding the bilocal operator around
coincident transverse points, $x_\perp=y_\perp$, since the kernel \eqref{delta}
localizes the transverse separation at $\Delta_\perp=0$ in the high-energy limit.

To study the high-energy/small-x$_B$ evolution, we write result \eqref{quarkSubEik-1pointinSWa} as 
\begin{eqnarray}
&&{1\over 2\pi\delta(0)}\int \dhd^4k\dbar(k^2)\theta(k^+)
\Big|\langle q(k)|\gamma^*_T(q)\rangle_{\small{\rm Fig.}\ref{Fig:DIS-q-1pointSW}a}\Big|^2
\nonumber\\
=\!\!\!&&\half\sum_f e^2 e_f^2\int d^2x\,{2\over sg^2}\Big(Q_1^f(x_\perp,x_B) + \bar{Q}_1^f(x_\perp,x_B)\Big)
+ O(\lambda^{-2}),
\label{quarkSubEik-Q1barQ1}
\end{eqnarray}
with
\begin{eqnarray}
&&\hspace{-1cm}Q_1^f(x_\perp,x_B) \equiv 
g^2{s\over 2}\!\!\int_{-\infty}^{+\infty}\!\! dx^+ \!\!\int_{-\infty}^{x^+}\!\! dy^+ \,e^{ix_B P^-\Delta^+}
\barpsi_f(y^+,x_\perp)[y^+,x^+]_x\ssn_1\psi_f(x^+,x_\perp)\,,
\label{Q1}
\end{eqnarray}
and 
\begin{eqnarray}
&&\hspace{-1cm}\bar{Q}_1^f(x_\perp,x_B) \equiv 
g^2{s\over 2}\!\!\int_{-\infty}^{+\infty}\!\! dy^+ \!\!\int_{-\infty}^{y^+}\!\! dx^+ \,e^{ix_B P^-\Delta^+}
\,\barpsi_f(y^+,x_\perp)[y^+,x^+]_x\ssn_1\psi_f(x^+,x_\perp)\,.
\label{Q1dagger}
\end{eqnarray}
Operators (\ref{Q1}), and (\ref{Q1dagger}) are the ones we found in
ref.~\cite{Chirilli:2021lif} with $x_B=0$. 

Adding also the contribution for diagram in Fig. \ref{Fig:DIS-q-1pointSW}b, we have
\begin{eqnarray}
&&\calm_{\rm Quark}^{T,LO} \equiv {1\over 2\pi\delta(0)}\int \dhd^4k\dbar(k^2)\theta(k^+)
\Big(\Big|\langle q(k)|\gamma^*_T(q)\rangle_{\small{\rm Fig.}\ref{Fig:DIS-q-1pointSW}a}\Big|^2
+ \Big|\langle \barq(k)|\gamma^*_T(q)\rangle_{\small{\rm Fig.}\ref{Fig:DIS-q-1pointSW}b}\Big|^2
\Big)
\nonumber\\
=\!\!\!&&i\sum_f e^2 e_f^2\int d^2x\,dx^+ dy^+\,\sin\big({x_B P^-\Delta^+}\big)
\bar\psi_f(y^+,x_\perp)\,[y^+,x^+]_x\,\ssn_1\,\psi_f(x^+,x_\perp)
+ O(\lambda^{-2}),
\label{qantquarkSubEik-1pointinSW}
\end{eqnarray}

Considering the reduced matrix element of \eqref{Int-squMEq-1pointSW-Ta}, from \eqref{qantquarkSubEik-1pointinSW} we obtain
($\alpha_{\rm em }={e^2\over 4\pi}$ and $\alpha_s = {g^2\over 4\pi}$)
\begin{eqnarray}
\langle\langle\calm_{\rm Quark}^{T,LO}\rangle\rangle = 
8\pi^2\alpha_{\rm em }\sum_fe^2_f
\Big(q_f(x_B) + \barq_f(x_B)\Big) + O(\lambda^{-2})\,.
\label{reduceMEquarkT}
\end{eqnarray}
with
\begin{eqnarray}
&&\hspace{-1cm}q_f(x_B) = {1\over 4\pi}\int_{-\infty}^{+\infty}\!\! dx^+\,e^{ix_BP^-x^+}
\bra{N(P)}\hat{\barpsi}_f(0^+,0_\perp)\gamma^-[0^+,x^+]\hat{\psi}_f(x^+,0_\perp)\ket{N(P)}
\label{qf}
\\
&&\hspace{-1cm}\barq_f(x_B) =  - {1\over 4\pi}\int_{-\infty}^{+\infty}\!\! dx^+ \,e^{ix_BP^-x^+}
\bra{N(P)}\hat{\barpsi}_f(x^+,0_\perp)\gamma^-[x^+,0^+]\hat{\psi}_f(0^+,0_\perp)\ket{N(P)}
\label{barqf}
\end{eqnarray}
The distributions (\ref{qf}), and (\ref{barqf}) are the quark and antiquark pdf, respectively, at $x_B\ne 0$.

We can now calculate the cross-section using \eqref{reduceMEquarkT}
\begin{eqnarray}
\sigma^{\gamma^*P}_T =\!\!\!&& {1\over 4P\cdot q}\langle\langle\calm_{\rm Quark}^{T,LO}\rangle\rangle
\nonumber\\
=\!\!\!&& {4\pi^2\over s}\alpha_{\rm em}\sum_f e^2_f\Big(q_f(x_B) + \barq_f(x_B)\Big)
\end{eqnarray}

\subsection{Helicity distributions from asymmetry}

We now repeat the same analysis for the asymmetry, which gives access to the helicity
distribution. As in the unpolarized case, one may first consider the formal $x_B=0$ limit.
However, the proper small-$x_B$ helicity distribution is obtained only after completing the
phase-space integration and only afterwards taking the small-$x_B$ limit.

For this we have to take the asymmetric combination of the virtual photon polarization
$(\varepsilon^i_+\varepsilon^{*j}_+ - \varepsilon^i_-\varepsilon^{*j}_-) = i\epsilon^{ji}$, thus the matrix element is

\begin{eqnarray}
&& i \epsilon^{ik}\barpsi\Big((g^{ij} + i\epsilon^{ij}\gamma^5)\ssn_2k_j + k^+(\gamma^i+i\gamma_j\gamma^5\epsilon^{ij})\Big)\ssk
\Big(\ssn_2(g^{kl}+i\epsilon^{kl}\gamma^5)k_l + k^+(\gamma^k-i\epsilon^{kl}\gamma_l\gamma^5)\Big)\psi
\nonumber\\
&& = - 8\,(k^+)^3\barpsi\gamma^5\ssn_1\psi + O(\lambda^{-2})
\end{eqnarray}
and the scattering amplitude square is
\begin{eqnarray}
&&\hspace{-1.4cm}{1\over 2\pi\delta(0)}\int \dhd^4k\dbar(k^2)\Big|\langle q(k)|\gamma^*_A(q)\rangle_{\small{\rm Fig.}\ref{Fig:DIS-q-1pointSW}a}\Big|^2\theta(k^+)
\nonumber\\
&&\hspace{-1.4cm}
= -\sum_fe^2e^2_f\,\int d^2x d^2y\dhd^2k e^{i(q-k,x-y)_\perp}\int dx^+ dy^+
\, e^{-i\left( q^- - {k^2_\perp\over 2q^+}\right)(x^+-y^+)}
\nonumber\\
&&\hspace{-1.4cm}
\times\big\langle\barpsi(y^+,y_\perp)[y^+,+\infty n_1]_y\gamma^5\ssn_1[+\infty n_1,x^+]_x\psi(x^+,x_\perp)\big\rangle +O(\lambda^{-2})
\label{squMEq-1pointSW-A}
\end{eqnarray}
We may now repeat the same steps we performed in the transverse polarization case in previous subsection. 
In eq. (\ref{squMEq-1pointSW-A}), if we set the exponential factors to $1$, and then perform the integration over $k_\perp$, 
we end up to the helicity distributions at $x_B=0$.

However, as we observed in the previous subsection, the proper way is to complete the phase space integration before taking the small-x limit. 
So, let us integrate over $k_\perp$, and from eq. (\ref{squMEq-1pointSW-A}), using \eqref{delta}, we obtain
\begin{eqnarray}
&&\hspace{-1.3cm}\calm^{A,LO}_{\rm Quark} \equiv {1\over 2\pi\delta(0)}\int \dhd^4k\dbar(k^2)\theta(k^+) 
\left(\Big|\langle q(k)|\gamma^*_A(q)\rangle_{\small {\rm Fig.}\ref{Fig:DIS-q-1pointSW}a}\Big|^2
+ \Big|\langle \barq(k)|\gamma^*_A(q)\rangle_{\small {\rm Fig.}\ref{Fig:DIS-q-1pointSW}b}\Big|^2\right)
\nonumber\\
&& = - 2\sum_f e^2e^2_f\int d^2xdx^+ dy^+ \,\cos\Big(x_BP^-\Delta^+\Big)
\big\langle\barpsi_f(y^+,x_\perp)[y^+,x^+]_x\gamma^5\ssn_1\psi_f(x^+,x_\perp)\big\rangle
\label{quarkSubEik-5pointinSW}
\end{eqnarray}
Considering, again, the reduced matrix element, from \eqref{quarkSubEik-5pointinSW} we obtain
\begin{eqnarray}
\sigma^{\gamma^* P}_A =\!\!\!&&
{1\over 4P\cdot q}\langle\langle\calm^{A,LO}_{\rm Quark}\rangle\rangle
\nonumber\\
=\!\!\! &&  {8\pi^2\over s}\alpha_{\rm em}\sum_f e^2_f 
\Big(\Delta q_f(x_B) + \Delta\barq_f(x_B)\Big) + O(\lambda^{-2})
\label{dipoXsec1pointSW-5}
\end{eqnarray}
with
\begin{eqnarray}
&&\hspace{-1.3cm}\Delta q_f(x_B) = - {1\over 4\pi}\int_{-\infty}^{+\infty}\!\! dx^+\,e^{ix_BP^-x^+}
\bra{N(P)}\hat{\barpsi}_f(0^+,0_\perp)\gamma^5\gamma^-[0^+,x^+]\hat{\psi}_f(x^+,0_\perp)\ket{N(P)}
\label{q5f}
\\
&&\hspace{-1.3cm}\Delta \barq_f(x_B) =  - {1\over 4\pi}\int_{-\infty}^{+\infty}\!\! dx^+ \,e^{ix_BP^-x^+}
\bra{N(P)}\hat{\barpsi}_f(x^+,0_\perp)\gamma^5\gamma^-[x^+,0^+]\hat{\psi}_f(0^+,0_\perp)\ket{N(P)}
\label{barq5f}
\end{eqnarray}
As we did for the unpolarized case, to study the high-energy/small-$x_B$ evolution, it is convenient to introduce the operators
\begin{eqnarray}
&&\hspace{-1cm}Q_{5,f}(x_\perp,x_B) \equiv g^2{s\over 2}\!\!\int_{-\infty}^{+\infty}\!\! dx^+ \!\!\int_{-\infty}^{x^+}\!\! dy^+
\,e^{ix_BP^-\Delta^+}\barpsi_f(y^+,x_\perp)[y^+,x^+]_x
\gamma^5\ssn_1\psi_f(x^+,x_\perp)\,,
\label{Q5}
\end{eqnarray}
and
\begin{eqnarray}
&&\hspace{-1cm}\bar{Q}_{5,f}(x_\perp,x_B) \equiv
g^2{s\over 2}\!\!\int_{-\infty}^{+\infty}\!\! dy^+ \!\!\int_{-\infty}^{y^+}\!\! dx^+ \,e^{ix_BP^-\Delta^+}
\barpsi_f(y^+,x_\perp)[y^+,x^+]_x\gamma^5\ssn_1\psi_f(x^+,x_\perp)\,.
\label{Q5dagger}
\end{eqnarray}
and write the square matrix element as
\begin{eqnarray}
&&{1\over 2\pi\delta(0)}\int \dhd^4k\dbar(k^2)\theta(k^+) 
\Big|\langle q(k)|\gamma^*_A(q)\rangle_{\small {\rm Fig.}\ref{Fig:DIS-q-1pointSW}a}\Big|^2
\nonumber\\
=&& -\sum_f e^2e^2_f\int d^2x\,{2\over sg^2}\Big(Q_{5,f}(x_\perp,x_B)+ \bar{Q}_{5,f}(x_\perp,x_B)\Big)
\end{eqnarray}

In the next section, we derive the same results we obtained in this section using the collinear non-local OPE formalism with
high-energy limit.

\section{From non-local OPE to High-energy OPE}
\label{sec:nonOPE2heOPE}

\begin{figure}[t]
	\begin{center}
		\includegraphics[width=4in]{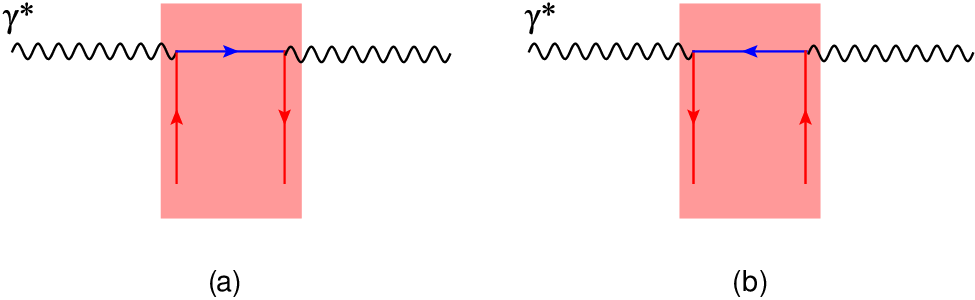}
		\caption{Diagrams contributing to the calculation of $T_{\mu\nu}$}
		\label{Fig:DIS-psipsibar}
	\end{center}
\end{figure}

In this section we show how to recover, through a different procedure, the same result
obtained in the previous section. We will do this only for the transverse photon polarization.
Our starting point is the non-local OPE developed by Balitsky and Braun in
ref.~\cite{Balitsky:1987bk,Balitsky:1990ck}, applied to the time-ordered product of two
electromagnetic currents.

The purpose of this section is twofold. First, it provides an independent derivation of the
operator structure entering the DIS cross section at sub-eikonal level. Second, it clarifies
how the operators appearing in the high-energy Wilson-line formalism emerge from the
high-energy limit of the non-local light-ray OPE. In this way, the connection between the
two formalisms becomes explicit.

Let us stress that the non-local OPE used in this section is the leading
light-cone, or leading-twist, approximation to the current-current product
in the Bjorken limit,
\begin{eqnarray}
	Q^2\to\infty,\qquad x_B={Q^2\over 2P\cdot q}\quad {\rm fixed}\,.
\end{eqnarray}
Thus, the expansion is performed first near the light cone, keeping the
leading-power nonlocal quark operator.  Only after this leading-power
non-local OPE has been obtained do we take its high-energy limit.  The
purpose of this section is therefore to show that
the leading-twist non-local light-ray operator, when subsequently taken to
the high-energy limit, reproduces the same inclusive quark operator content
obtained from the first sub-eikonal correction in the shock-wave
calculation. Thus, the comparison performed here is a
leading-power comparison in the Bjorken expansion. It is not meant to imply
a general one-to-one correspondence between all higher-twist terms and all
sub-eikonal corrections.

We start from the forward Compton amplitude
\begin{eqnarray}
T_{\mu\nu} = i\int d^4x \, e^{iq\cdot x}\bra{N(P)}{\rm T}\{\hat{j}_\mu(x)\hat{j}_\nu(0)\}\ket{N(P)}\,.
\label{Tmunu}
\end{eqnarray}
Our goal is to calculate $T_{\mu\nu}$ at leading order in the background-field method.
To this end, we separate all fields into quantum and classical parts. At LO, one has two
diagrams (see Fig.~\ref{Fig:DIS-psipsibar}),
\begin{eqnarray}
	&&\hspace{-0.6cm} T_{\mu\nu} =  i\int d^4x d^4y \, e^{iq\cdot (x-y)}
	{i\over 2\pi^2[(x-y)^2-i\epsilon]^2}
	\nonumber\\
	&&\hspace{0.6cm}\times
	\Big(\big\langle{\rm T}\{\barpsi_c(x)\gamma_\mu(\ssx-\ssy) [x,y] \gamma_\nu\psi_c(y)\}\big\rangle_A
	\nonumber\\
	&&\hspace{0.9cm}
	-\big\langle{\rm T}\{\barpsi_c(y)\gamma_\nu(\ssx-\ssy)[y,x]\gamma_\mu\psi_c(x)\}\big\rangle_A\Big)\,.
	\label{LO-Tmunu-a}
\end{eqnarray}
Here we used translation invariance of the matrix element;
the subscript $c$, which denotes classical fields, will be omitted in the following, and the subscript $A$ means, as usual, matrix element in the 
background of the gluon field.

To obtain Eq.~\eqref{LO-Tmunu-a}, we used the leading light-cone term of the quark propagator in the background gluon field, before taking the high-energy limit, given by~\cite{Balitsky:1987bk}
\begin{eqnarray} 
\langle{\rm T}\{\psi_q(x)\barpsi_q(y)\}\rangle_A 
= {i(\ssx-\ssy)\over 2\pi^2[(x-y)^2-i\epsilon]^2}[x,y]_c + O((x-y)^{-2})\,.
\label{lightc-expansion}
\end{eqnarray}
where the subscript $q$ denotes quantum fields, and $[x,y]_c$ is the straight gauge link made of
the classical gluon field connecting the two points $x^\mu$ and $y^\mu$ and defined as
\begin{eqnarray}
[x,y]_c= \mathrm P\exp\!\left\{ig\int_y^x dz_\mu\,A_c^\mu(z)\right\}\,.
\label{gaugelink}
\end{eqnarray}
The term explicitly kept in Eq.~\eqref{lightc-expansion} is the leading light-cone
singularity and gives the leading-twist non-local quark operator in the Bjorken
expansion. The terms denoted by \(O((x-y)^{-2})\) are less singular on the light cone
than the term explicitly kept in Eq.~\eqref{lightc-expansion}. They correspond to
higher-twist, power-suppressed contributions in the Bjorken expansion and
are not needed for the leading-power comparison performed in this section.

This is possible because the Bjorken leading-power expansion and the
high-energy eikonal expansion are different power countings: an operator
which is leading twist in the collinear light-cone OPE may correspond to a
sub-eikonal contribution when reorganized in the high-energy shock-wave
power counting.

The equation \eqref{LO-Tmunu-a} is written in terms of a straight gauge link connecting the two
space-time points of the non-local quark operator. At this stage, no high-energy
approximation has yet been made. The point is that, after taking the high-energy limit, this
straight gauge link reorganizes into the Wilson-line structures which are natural in the
shock-wave formalism. This is precisely the mechanism by which the non-local OPE is
mapped into the high-energy OPE.

Under the infinite longitudinal boost we are considering, it is not difficult to show that, 
the gauge link $[x,y]$ reduces to the light-cone gauge link at fixed transverse position (see Appendix \ref{sec:straightgaugelink})
\begin{eqnarray}
[x,y] \stackrel{\lambda\to \infty}{\longrightarrow} [x^+n_1+x_\perp, y^+n_1+x_\perp]\,.
\end{eqnarray}

The classical spinor field is also modified by the high-longitudinal boost by becoming independent of the $x^-$ component
$\psi(x)\to \psi(x^+,x_\perp)$.

With these approximations, the eq. (\ref{LO-Tmunu-a}) becomes
\begin{eqnarray}
\hspace{-1cm}T_{\mu\nu} =&&\!\!\! i\int d^4x d^4y \, e^{iq\cdot (x-y)}
{i\over 2\pi^2[(x-y)^2-i\epsilon]^2}\sum_fe^2_f
\nonumber\\
&&\hspace{0.6cm}\times
\Big(\big\langle{\rm T}\{\barpsi_c(x^+,x_\perp)\gamma_\mu(\ssx-\ssy) [x^+,y^+]_x \gamma_\nu\psi_c(y^+,y_\perp)\}\big\rangle_A
\nonumber\\
&&\hspace{0.9cm}
-\big\langle{\rm T}\{\barpsi_c(y^+,y_\perp)\gamma_\nu(\ssx-\ssy) [y^+,x^+]_y\gamma_\mu\psi_c(x^+,x_\perp)\}\big\rangle_A\Big)
\label{non-localOPE}
\end{eqnarray}

Since the longitudinal polarization will give a sub-leading contribution, 
let us consider the transverse polarization only, and use the approximations given in the eq. (\ref{spinorboost}). 
The same exact procedure can be applied to obtain the result (\ref{squMEq-1pointSW-A}) for the asymmetric combination of the virtual photon polarization.
Thus, using
\begin{eqnarray}
\barpsi\gamma^i(\ssx-\ssy)\gamma^i\psi = 2(x^+-y^+)\barpsi\ssn_1\psi + O(\lambda^{-2})
\end{eqnarray}
from (\ref{non-localOPE}), we have ($i,j=1,2$ as usual)
\begin{eqnarray}
\nonumber\\
&&\half\sum_{\lambda=\pm 1}\varepsilon^i_\lambda\varepsilon^{*j}_\lambda T_{ij}
\nonumber\\
=\!\!\!&& - \lim_{\epsilon^+\to 0} 
\sum_fe^2_f\!\int dx^+dx^- d^2x dy^+dy^- d^2y \,
{\Delta^+\over 4\pi^2}\,{e^{iq^+\Delta^- + iq^-\Delta^+ - i(q,\Delta)_\perp} \over \left[2\Delta^-\Delta^+ - \Delta_\perp^2 - i\epsilon\right]^2}
\nonumber\\
&&\times \Big(\bra{N(P)}\hat{\barpsi}(x^+,x_\perp)\ssn_1[x^+,y^+]_x\hat{\psi}(y^+,y_\perp)\ket{N(P+\epsilon^+ n_1)}
\nonumber\\
&&- \bra{N(P)}\hat{\barpsi}(y^+,y_\perp)\ssn_1[y^+,x^+]_y\hat{\psi}(x^+,x_\perp)\ket{N(P+\epsilon^+ n_1)}
+O(\lambda^{-2})
\end{eqnarray}
with $\Delta^\mu=(x-y)^\mu$.

Let us observe that the reduced matrix element is defined as
\begin{eqnarray}
\langle\!\langle \mathcal O \rangle\!\rangle
\equiv
\lim_{\epsilon^+\to 0}
\frac{\bra{N(P)}\mathcal O\ket{N(P+\epsilon^+ n_1)}}
{\dbar(\epsilon^+)\,\dbar(0)\,\dbar^{(2)}(0)} \,.
\label{reducedME2}
\end{eqnarray}
In this case, the reduced matrix element is obtained by dividing out the overall volume factors associated 
with the average position of the bilocal operator. Writing the operator in terms of the average and relative coordinates,
\begin{eqnarray}
X^\mu=\frac{x^\mu+y^\mu}{2},
\qquad
\Delta^\mu=x^\mu-y^\mu\,,
\end{eqnarray}
one sees that, for a forward matrix element, the dependence on the center coordinate $X^\mu=(X^+,X^-,X_\perp)$ drops out, 
so that the integrations over $X^\mu$ factorize. The factor $\dbar^{(2)}(0)$ comes from the integration over the transverse 
impact parameter $X_\perp$, i.e. from averaging over the transverse position of the projectile with respect to the shock wave. 
The factor $\dbar(\epsilon^+)$, which becomes $\dbar(0)$ in the forward limit, is associated with the invariance under a common shift
of the two operator insertions along the $x^+$ direction. Finally, the extra factor $\dbar(0)$ arises from the integration over the average $X^-$ coordinate, reflecting the invariance under translations of the bilocal operator along the $x^-$ direction. Thus the reduced matrix element is defined after removing the overall volume factor associated with the three center-of-mass coordinates $X^+$, $X^-$, and $X_\perp$.

Using $q^-= -{Q^2\over 2q^+}=-x_BP^-$ at $q_\perp=0$, and taking the residue integrating over $\Delta^-$ we obtain
\begin{eqnarray}
\hspace{-1cm}\half\!\!\sum_{\lambda=\pm 1}\varepsilon^i_\lambda\varepsilon^{*j}_\lambda T_{ij} 
&&= \lim_{\epsilon^+\to 0} \half \sum_fe^2_f\int dx^+dy^+\theta(x^+-y^+)d^2x d^2y 
\,{q^+\over 2\pi\Delta^+} e^{i q^+{\Delta^2_\perp\over 2\Delta^+} - i{Q^2\over 2q^+}\Delta^+ }
\nonumber\\
&&\times \Big(\bra{N(P)}\hat{\barpsi}(x^+,x_\perp)\ssn_1[x^+,y^+]_x\hat{\psi}(y^+,y_\perp)\ket{N(P+\epsilon^+ n_1)}
\nonumber\\
&&	- \bra{N(P)}\hat{\barpsi}(y^+,y_\perp)\ssn_1[y^+,x^+]_y\hat{\psi}(x^+,x_\perp)\ket{N(P+\epsilon^+ n_1)}\Big) + O(\lambda^{-2})	
\end{eqnarray}
Following the procedure used in the previous section, we can now take the high-energy limit, and using 
${-iq^+\over 2\pi \Delta^+}\exp\left\{iq^+{\Delta^2_\perp\over 2\Delta^+}\right\} = \delta^{(2)}(\Delta_\perp)$ we arrive at
\begin{eqnarray}
\hspace{-1cm}\half\!\!\sum_{\lambda=\pm 1}\varepsilon^i_\lambda\varepsilon^{*j}_\lambda T_{ij} 
&&= {i\over 2}\lim_{\epsilon^+\to 0}\sum_fe^2_f\int dx^+dy^+d^2x\,\theta(x^+-y^+)\,e^{- i{Q^2\over 2q^+}\Delta^+ }
\nonumber\\
&&\times \Big(\bra{N(P)}\hat{\barpsi}(x^+,x_\perp)\ssn_1[x^+,y^+]_x\hat{\psi}(y^+,x_\perp)\ket{N(P+\epsilon^+ n_1)}
\nonumber\\
&&	- \bra{N(P)}\hat{\barpsi}(y^+,x_\perp)\ssn_1[y^+,x^+]_x\hat{\psi}(x^+,x_\perp)\ket{N(P+\epsilon^+ n_1)}\Big) + O(\lambda^{-2})	
\end{eqnarray}
Using translation invariance and the reduced matrix elements, we have
\begin{eqnarray}
\hspace{-1cm}\half\!\!\sum_{\lambda=\pm 1}\varepsilon^i_\lambda\varepsilon^{*j}_\lambda T_{ij} 
=\sum_fe^2_f\int_0^{+\infty}\!\! dx^+\,e^{- i{Q^2\over 2q^+}x^+ }
{\rm Im}\bra{N(P)}\hat{\barpsi}(0^+,x_\perp)\ssn_1[0^+,x^+]_x\hat{\psi}(x^+,x_\perp)\ket{N(P)} + O(\lambda^{-2})	
\label{Tij}
\end{eqnarray}
We are now ready to take the imaginary part of eq. \eqref{Tij}
\begin{eqnarray}
&&\half\!\!\sum_{\lambda = \pm 1}\varepsilon^i_\lambda(q)\varepsilon^{*j}_\lambda(q) {\rm Im}T_{ij}
\nonumber\\
= && -\sum_fe^2_f\int_0^{+\infty}\!\! dx^+\,\sin\Big(x_BP^-x^+ \Big)
{\rm Im}\bra{N(P)}\hat{\barpsi}(0^+,x_\perp)\ssn_1[0^+,x^+]_x\hat{\psi}(x^+,x_\perp)\ket{N(P)} + O(\lambda^{-2})	
\nonumber\\
= && \pi \sum_fe^2_f\Big(q_f(x_B) + \barq_f(x_B)\Big) + O(\lambda^{-2})	
\label{Tmunu-qbaqrq}
\end{eqnarray}
where $q_f$ and $\barq_f$ are defined in (\ref{qf}) and (\ref{barqf}), respectively, and where we used ${Q^2\over 2q^+} = x_BP^-$

We can now compare the two derivations using (the factor $4\pi\alpha_{\rm em}$ eliminates the electric charge $e^2$)
\begin{eqnarray}
\hspace{-1cm}\varepsilon^\mu\varepsilon^{*\nu}{1\over 2\pi}{\rm Im}\,T_{\mu\nu} 
=\!\!\!&& \varepsilon^\mu\varepsilon^{*\nu}W_{\mu\nu} 
\nonumber\\
=\!\!\!&& {1\over 4\pi}{1\over 4\pi\alpha_{\rm em}}\langle\langle\calm^{T,LO}_{\rm Quark}\rangle\rangle
\label{comparisonWTmunu}
\end{eqnarray}
From (\ref{comparisonWTmunu}), and using result (\ref{reduceMEquarkT}), and (\ref{Tmunu-qbaqrq}), 
we deduce that, 
\begin{eqnarray}
&&\hspace{-1.1cm}\half\!\!\sum_{\lambda=\pm 1}\varepsilon^i_\lambda(q)\varepsilon^{*j}_\lambda(q)W_{ij}
= {1\over  4\pi}{1\over 4\pi\alpha_{\rm em}}\Big[8\pi^2\alpha_{\rm em}\sum_fe^2_f(q_f(x_B)+\barq_f(x_B))\Big]
\\
&&\hspace{-1.1cm} \half\!\!\sum_{\lambda=\pm 1}\varepsilon^i_\lambda(q)\varepsilon^{*j}_\lambda(q){1\over 2\pi}{\rm Im}\,T_{ij}
= {1\over 2\pi}\,\Big[\pi
\sum_fe^2_f\big(q_f(x_B) + \barq_f(x_B)\big) \Big]
\label{compare}
\end{eqnarray}

The result, eq.~\eqref{compare}, shows that the connection between the high-energy shock-wave description of DIS and the familiar quark picture at finite 
Bjorken $x_B$ emerges already at first sub-eikonal order. Although the calculation is performed in the dipole shock-wave formalism, the inclusive cross 
section is governed not by a purely $x_B=0$ operator structure, but by the standard nonlocal quark and antiquark light-cone distributions $q_f(x_B)$ and 
$\bar q_f(x_B)$. In other words, once the high-energy limit is taken only in the transverse kernel, while the longitudinal phase $e^{ix_B P^- \Delta^+}$ is 
kept intact, the sub-eikonal formalism reproduces the correct longitudinal operator content needed for the partonic interpretation of DIS at finite $x_B$. The 
agreement of the two derivations therefore provides an explicit operator-level bridge between the dipole shock-wave formalism and the conventional 
light-cone description.

\section{Evolution equation}
\label{sec:evolutionQ1Q5}

\begin{figure}[thb]
	\begin{center}
		\includegraphics[width=4.0in]{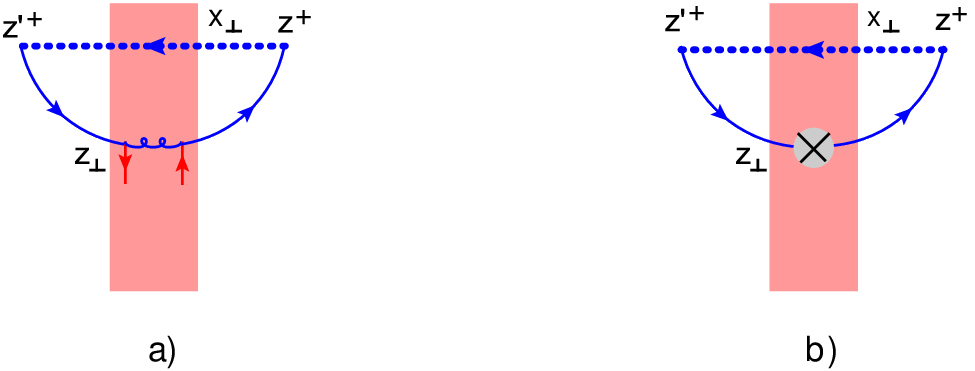}
		\caption{Diagrams with $\hat{Q}_{1x}$ and $\hat{Q}_{5x}$ quantum. In diagram in panel a) we have quarks in the background, while in diagram
		in panel b) we have gluon sub-eikonal corrections denoted by the crossed gray circle.}
		\label{LO-psibarpsiadj4}
	\end{center}
\end{figure}

The energy dependence of the cross section derived above, written as a convolution of
coefficient functions and matrix elements of operators, reflects the semiclassical nature of
the high-energy formalism. According to the high-energy OPE~\cite{Balitsky:1995ub}, the
evolution of these operators resums the large logarithms in rapidity and restores unitarity
of the theory. At eikonal level this is precisely what happens for Wilson lines, whose
evolution is governed by the BK/B-JIMWLK equations~\cite{Balitsky:1995ub, Kovchegov:1999yj, Kovchegov:1999ua, Jalilian-Marian:1997qno, 
Jalilian-Marian:1997jhx, Jalilian-Marian:1997ubg, Weigert:2000gi, Iancu:2000hn,
Ferreiro:2001qy, Gelis:2010nm} (see ref.~\cite{Balitsky:2001gj, Balitsky:2025bup, Kovchegov:2012mbw,Gelis:2010nm} for review).

At sub-eikonal level the situation is similar. The cross section is now convoluted with new
operators, and in order to determine its energy dependence one has to derive the evolution
equations for these operators. In the approximation in which the fields are ordered
according to the Sudakov component $p^+$, the evolution equations of the operators 
$Q^f_1(x_\perp)$ and $Q^f_5(x_\perp)$, eqs. \eqref{Q1}, and \eqref{Q5}, were calculated in ref.~\cite{Chirilli:2021lif}, 
and we report them here. Examples of diagrams contributing to the evolution of these operators are shown in figure \ref{LO-psibarpsiadj4}.
We will use notation $\tr$ for trace over spinor indices, and $\Tr$ for trace over color indices in the fundamental representation.

The evolution equations for $Q^f_1(x_\perp)$ and  $Q^f_5(x_\perp)$ are\footnote{Note that here the operators $Q_1(x_\perp)$ and $Q_5(x_\perp)$
have an overall sign difference with respect to the ones defined in ref~\cite{Chirilli:2021lif}.}
(the evolution equations for the Hermitian-conjugate operators are obtained analogously)
\begin{eqnarray}
\hspace{-2cm}{d\over d\eta} Q^f_{1\,x}
=\!\!\!&& {\alpha_s\over 4\pi^2}\int \,
{d^2z\over (x-z)^2_\perp}\bigg( Q^f_{1\,z}{\rm Tr}\{U_z U^\dagger_x \}
+ {1\over N_c}{\rm Tr}\{U^\dagger_x \tilde{Q}^f_{1\,z}\} \bigg)
\label{evolutionQ1a}
\end{eqnarray}
and
\begin{eqnarray}
{d\over d\eta}Q^f_{5x}
={\alpha_s\over 4\pi^2}\int  {d^2z \over (x-z)^2_\perp}\Big[\Tr\{U^\dagger_xU_z\}Q^f_{5\,z} 
+ {1\over N_c}\Tr\{U^\dagger_x\tildeQ^f_{5z}\} - 2\Tr\{U^\dagger_x \calf_z\} \Big]
\label{evolutionQ5a}
\end{eqnarray}
with superscript $f$ the flavor index, and where we defined
\begin{eqnarray}
&&\calf_z = \calf(z_\perp) \equiv ig{s\over 2}\int_{-\infty}^{+\infty}\!\!dz^+[\infty n_1,z^+]_z
\,\epsilon^{ij}F_{ij}\big(z^+,z_\perp\big)[z^+,-\infty n_1]_z\,.
\label{calfoperator_a}
\end{eqnarray}
and
\begin{eqnarray}
\hspace{-1.3cm}\tildeQ^f_{1ij}(x_\perp) \equiv\!\!\!&& g^2{s\over 2}\!\!\int_{-\infty}^{+\infty}\!\!dz^+\!\!\int_{-\infty}^{z^+}\!\!dz'^+
\big([\infty n_1,z^+]_x {\rm tr}\{\psi_f(z^+,x_\perp)\bar{\psi}_f(z'^+,x_\perp)\,\ssn_1\}[z'^+, -\infty n_1]_x\big)_{ij}\,,
\label{Qt1}
\\
\tildeQ^f_{5ij}(x_\perp) \equiv\!\!\!&& g^2{s\over 2}\!\!\int_{-\infty}^{\infty}dz^+\!\int_{-\infty}^{z^+}\!dz'^+
\big([\infty n_1, z^+]_x\tr\{\psi_f(z^+,x_\perp)\barpsi_f(z'^+,x_\perp)\gamma^5\ssn_1\}[z'^+,-\infty n_1]_x\big)_{ij}
\label{Qt5}
\end{eqnarray}
where $i, j$ are color indexes in the fundamental representation.

The parameter $\eta$ is the rapidity parameter which enters as a rigid cut-off of the logarithmic divergence in the longitudinal momentum $k^+$.
Evolution equations \eqref{evolutionQ1a}, and \eqref{evolutionQ5a} can be rewritten as follows (the Hermitian-conjugate equations are obtained analogously)
\begin{eqnarray}
\hspace{-2cm}{d\over d\eta}Q^{\rm S}_{1\,x}
={\alpha_s\over 4\pi^2}\int \,
{d^2z\over (x-z)^2_\perp}\bigg(2C_F Q^{\rm S}_{1z} - N_c\calq^{\rm S}_{1zx} - {1\over N_c}\Psi^{\rm S}_{1zx} \bigg)
\label{evolutionQ1b}
\end{eqnarray}
\begin{eqnarray}
\hspace{-2cm}{d\over d\eta}Q^{\rm S}_{5\,x}
={\alpha_s\over 4\pi^2}\int \,
{d^2z\over (x-z)^2_\perp}\bigg(2C_F Q^{\rm S}_{5z} - N_c\calq^{\rm S}_{5zx} - {1\over N_c}\Psi^{\rm S}_{5zx} - 2N_f\calf_{xz}\bigg)
\label{evolutionQ5b}
\end{eqnarray}
($C_F={N_c^2-1\over 2N_c}$) with the flavor singlet operators defined as
\begin{eqnarray}
&&Q^{\rm S}_{1\,x} = \sum_{f=1}^{N_f}Q^f_{1\,x}\,, \qquad 
\Psi^{\rm S}_{1\,xz} = \sum_{f=1}^{N_f} \Psi^f_{1\,xz}\,, \\
&&Q^{\rm S}_{5\,x} = \sum_{f=1}^{N_f}Q^f_{5\,x}\,, \qquad \calq^{\rm S}_{5\,zx} = \sum_{f=1}^{N_f} \calq^f_{5\,zx}\,,
\qquad \Psi^{\rm S}_{5\,xz} = \sum_{f=1}^{N_f} \Psi^f_{5\,xz}\,,
\end{eqnarray}
and 
\begin{eqnarray}
&&\hspace{-2cm}{d\over d\eta}Q^{{\rm NS},(a)}_{1\,x}
={\alpha_s\over 4\pi^2}\int \,{d^2z\over (x-z)^2_\perp}\bigg(2C_F Q^{{\rm NS},(a)}_{1z} - N_c\calq^{{\rm NS},(a)}_{1zx} 
- {1\over N_c}\Psi^{{\rm NS},(a)}_{1zx}\bigg)
\label{evolutionQ1NSb}
\\
&&\hspace{-2cm}{d\over d\eta}Q^{{\rm NS},(a)}_{5\,x}
={\alpha_s\over 4\pi^2}\int \,{d^2z\over (x-z)^2_\perp}\bigg(2C_F Q^{{\rm NS},(a)}_{5z} - N_c\calq^{{\rm NS},(a)}_{5zx} 
- {1\over N_c}\Psi^{{\rm NS},(a)}_{5zx}\bigg)
\label{evolutionQ5NSb}
\end{eqnarray}
with the flavor non-singlet operators defined as 
\begin{eqnarray}
Q^{{\rm NS},(a)}_{1,5\,x} = \sum_{f=1}^{N_f} c_f^{(a)}Q^f_{1,5\,x} \,, \qquad \sum_{f=1}^{N_f}c_f^{(a)} = 0\,,
\end{eqnarray}
where the vector $c_f^{(a)} = (c_1^{(a)},\dots, c_{N_f}^{(a)})$ selects one non-singlet combination.
In eqs. \eqref{evolutionQ1b}, \eqref{evolutionQ5b}, and \eqref{evolutionQ5NSb},  we introduced the dipole-type operators 
\begin{eqnarray}
&&\calq^f_{1\,xy} = Q^f_1(x_\perp,y_\perp) \equiv Q^f_{1\,x}\calu_{xy}
\label{Q1xy}
\\
&&\calq^f_{5\,xy} = Q^f_5(x_\perp,y_\perp) \equiv Q^f_{5\,x}\calu_{xy}
\label{Q5xy}
\\
\nonumber\\
&&\Psi^f_{1\,xy} = \Psi^f_1(x_\perp,y_\perp) \equiv \Tr\{\tildeQ^f_{1x}\big(U^\dagger_x-U^\dagger_y\big)\}
\label{Psi1xy}
\\
&&\Psi^f_{5\,xy} = \Psi^f_5(x_\perp,y_\perp) \equiv \Tr\{\tildeQ^f_{5x}\big(U^\dagger_x-U^\dagger_y\big)\}
\label{Psi5xy}
\\
\nonumber\\
&&\calf_{xy} = \calf(x_\perp,y_\perp) \equiv \Tr\{U^\dagger_x\calf_y\}
\label{calFxy}
\end{eqnarray}
and the known dipole Wilson line operator
\begin{eqnarray}
\calu_{xy} =\calu(x_\perp,y_\perp) \equiv 1 - {1\over N_c}\Tr\{U(x_\perp)U^\dagger(y_\perp)\}\,.
\label{dipoleWilsonline}
\end{eqnarray}

Here the singlet/non-singlet decomposition refers to the flavor structure of the
$x_B=0$ high-energy operators and their evolution equations. It should not be
identified directly with the decomposition of the physical one-photon helicity
structure function, whose finite-$x_B$ contribution is weighted by the
electromagnetic charges $e_f^2$.

Note that the operators \eqref{Q1xy}-\eqref{calFxy} have the same property as the operator $\calu_{xy}$, namely they vanish for $x\to y$.
This makes manifest the dipole-type operator structures entering the leading-logarithmic high-energy corrections.
Indeed, as shown in ref.~\cite{Chirilli:2026vij}, the leading-logarithmic part of eqs.~\eqref{evolutionQ1b}, \eqref{evolutionQ5b}, and \eqref{evolutionQ5NSb} resums
the logarithms of energy that appear in the sub-eikonal corrections to the structure function $g_1$.

Notice also that, in the singlet case, eq.~\eqref{evolutionQ5b} contains mixing between the singlet quark sector
and the gluon operator $\calf_{xy}$, whereas in the non-singlet case such mixing is absent.

The evolution equations \eqref{evolutionQ1b}, \eqref{evolutionQ5b}, and \eqref{evolutionQ5NSb} are not closed, since they involve additional operators 
whose evolution must also be specified.
To obtain a closed system, one also needs the evolution equations for the operators \eqref{Q1xy}, and \eqref{Q5xy}, \eqref{Psi1xy}, \eqref{Psi5xy},
and \eqref{calFxy},
which were derived in ref.~\cite{Chirilli:2022dzt}.
In this work, however, we will not consider the full problem of solving the closed system of evolution equations, and we leave it for future work.

To isolate a strict local ladder approximation 
from the evolution equations \eqref{evolutionQ1b}, \eqref{evolutionQ5b}, and \eqref{evolutionQ5NSb}, one may project the evolution equations
onto the singular region $z_\perp\to x_\perp$, where the dipole-type combinations vanish
in the zero-dipole-size limit. In this reduced approximation, one retains only the local
terms proportional to $Q^f_{1z}$ and $Q^f_{5z}$, obtaining
\begin{eqnarray}
\hspace{-2cm}{d\over d\eta} Q^{{\rm NS}, (a)}_{1\,x}
=\!\!\!&& {\alpha_s\over 2\pi^2}C_F\!\!\int \!\! d^2z\,
{Q^{{\rm NS}, (a)}_{1z} \over (x-z)^2_\perp}
\label{doublelogQ1}
\end{eqnarray}
\begin{eqnarray}
\hspace{-2cm}{d\over d\eta} Q^{{\rm NS}, (a)}_{5\,x}
=\!\!\!&& {\alpha_s\over 2\pi^2}C_F\!\!\int \!\! d^2z\,
{Q^{{\rm NS}, (a)}_{5z} \over (x-z)^2_\perp}\,.
\label{doublelogQ5}
\end{eqnarray}
These equations define the reduced strict-ladder double-logarithmic approximation for
$Q_1^{\rm NS}$ and, in the non-singlet channel, for $Q_5^{NS,(a)}$. They should not be confused
with the full double-logarithmic behavior of the enlarged operator sector, whose complete
evolution is not closed in the present formulation.
The difference between the unpolarized and polarized sectors (non-singlet) is therefore not 
in the structure of the reduced ladder equation itself, but in the operator on which it acts 
and in the boundary conditions. The problem is thus reduced to solving a common double-logarithmic evolution equation. In the next subsection we analyze its solution and 
clarify how two distinct regimes emerge, depending on whether the transverse phase space is treated as independent of the longitudinal evolution variable or is instead 
constrained by the longitudinal ordering.

\subsection{Double-logarithmic solution of the $Q^{\rm NS}_1$ and $Q^{\rm NS}_5$ evolution}
\label{solution}

In the double-logarithmic approximation, the evolution of $Q^{\rm NS}_1$ (and similarly of $Q^{\rm NS}_5$) can be written as
\begin{eqnarray}
{d\over d\eta} Q^{\rm NS}_1(x_\perp,\eta) ={\alpha_s C_F \over 2\pi^2}
\int d^2 z\,
{Q^{\rm NS}_1(z_\perp,\eta)\over (x-z)_\perp^2}\,.
\label{Q1-G6-start}
\end{eqnarray}
The nature of the second logarithm depends on how the transverse phase space is treated. 
If the transverse integral is taken over an interval independent of the longitudinal variable, one obtains the usual mixed longitudinal-transverse double logarithm. 
By contrast, if the transverse phase space is constrained by the longitudinal ordering itself, the second logarithm becomes again a logarithm of energy. 
We discuss these two cases separately.

\subsubsection{Independent transverse phase space: BFKL/DGLAP double log}

To isolate the logarithm generated by the transverse integration, we introduce
\begin{eqnarray}
\rho \equiv \ln {L_\perp^2\over r_\perp^2},
\qquad r_\perp^2 \equiv (x-z)_\perp^2,
\qquad
0\le r_\perp^2 \le L_\perp^2\,.
\label{rho-def}
\end{eqnarray}
Here $L_\perp^2$ is a characteristic transverse size, treated for the moment as independent of the longitudinal evolution variable. 
Equation~\eqref{Q1-G6-start} then becomes (for simplicity, in what follows we will omit the superscript ${\rm NS}$)
\begin{eqnarray}
{\partial\over\partial\eta}Q_1(\eta,\rho) = \bar\alpha_s\int_0^\rho d\rho'\,Q_1(\eta,\rho'),
\qquad
\bar\alpha_s \equiv \frac{\alpha_s C_F}{2\pi},
\qquad
\eta\ge 0,\ \rho\ge 0.
\label{Q1-dla-integral}
\end{eqnarray}
Differentiating with respect to $\rho$, we obtain
\begin{eqnarray}
{\partial^2\over \partial\eta\,\partial\rho}Q_1(\eta,\rho) = \bar\alpha_s\,Q_1(\eta,\rho),
\qquad
\eta\ge 0,\ \rho\ge 0.
\label{Q1-dla-local}
\end{eqnarray}

To solve eq.~\eqref{Q1-dla-local}, we introduce the Green function
$G(\eta,\rho)$ defined by
\begin{eqnarray}
\left({\partial^2\over \partial\eta\,\partial\rho} - \bar\alpha_s \right)G(\eta,\rho) = \delta(\eta)\,\delta(\rho)\,,
\qquad
G(\eta,\rho)=0
\quad \text{for}\quad
\eta<0\ \text{or}\ \rho<0.
\label{Green-eq}
\end{eqnarray}
We perform the double Laplace transform
\begin{eqnarray}
\widetilde G(\omega,\gamma) \equiv \int_0^\infty d\eta\,e^{-\omega\eta}
\int_0^\infty d\rho\,e^{-\gamma\rho}\,G(\eta,\rho),
\qquad
\Re\,\omega>0,\ \Re\,\gamma>0, \qquad (\Re\,\omega)(\Re\,\gamma)> \bar\alpha_s
\label{Green-Laplace}
\end{eqnarray}
with the following limits from inside the positive quadrant
\begin{eqnarray}
\lim_{\rho\to 0^+}G(\eta,\rho)=1 
\qquad (\eta>0)\,,\qquad\qquad
\lim_{\eta\to 0^+}G(\eta,\rho)=1 
\qquad (\rho>0)\,.\qquad\qquad
\end{eqnarray}
Applying eq.~\eqref{Green-Laplace} to eq.~\eqref{Green-eq} gives
\begin{eqnarray}
(\omega\gamma-\bar\alpha_s)\,\widetilde G(\omega,\gamma)=1,
\end{eqnarray}
hence
\begin{eqnarray}
\widetilde G(\omega,\gamma) = {1\over \omega\gamma-\bar\alpha_s}.
\label{Green-transform}
\end{eqnarray}
The denominator in eq.~\eqref{Green-transform} yields the characteristic double-logarithmic relation
\begin{eqnarray}
\omega\gamma=\bar\alpha_s.
\label{dispersion-dla}
\end{eqnarray}

The inverse transform reads
\begin{eqnarray}
G(\eta,\rho) = \int_{c_\omega-i\infty}^{c_\omega+i\infty}\frac{d\omega}{2\pi i}
\int_{c_\gamma-i\infty}^{c_\gamma+i\infty}\frac{d\gamma}{2\pi i}\,
e^{\omega\eta+\gamma\rho}\,
{1\over \omega\gamma-\bar\alpha_s},
\qquad
c_\omega,c_\gamma>0\,, \qquad c_\omega c_\gamma>\bar\alpha_s\,.
\label{Green-inverse}
\end{eqnarray}
Performing first the $\omega$ integral gives
\begin{eqnarray}
G(\eta,\rho) = \theta(\eta)\int_{c_\gamma-i\infty}^{c_\gamma+i\infty}\frac{d\gamma}{2\pi i}\,
\frac{1}{\gamma}
\exp\!\left(\gamma\rho+\frac{\bar\alpha_s\,\eta}{\gamma}\right).
\label{Green-inverse-gamma}
\end{eqnarray}
Expanding the second exponential and using
\begin{eqnarray}
\int_{c_\gamma-i\infty}^{c_\gamma+i\infty}\frac{d\gamma}{2\pi i}\,
\gamma^{-n-1}e^{\gamma\rho}
=\theta(\rho)\,\frac{\rho^n}{n!},
\label{inverse-gamma-formula}
\end{eqnarray}
we obtain
\begin{eqnarray}
G(\eta,\rho) &=& \theta(\eta)\theta(\rho)
\sum_{n=0}^{\infty}
\frac{(\bar\alpha_s\,\eta\,\rho)^n}{(n!)^2}
\nonumber\\
&=&\theta(\eta)\theta(\rho)\,I_0\!\left(2\sqrt{\bar\alpha_s\,\eta\,\rho}\right),
\label{Green-final}
\end{eqnarray}
where $I_0$ is the modified Bessel function of the first kind.
For $\eta>0$ and $\rho>0$, the step functions in eq.~\eqref{Green-final}
are equal to unity. The physical solution $Q_1(\eta,\rho)$
is extended continuously to the axes.
For constant boundary conditions,
\begin{eqnarray}
Q_1(0,\rho)=Q_1^{(0)},
\qquad
Q_1(\eta,0)=Q_1^{(0)},
\label{Born-bc}
\end{eqnarray}
the solution of eq.~\eqref{Q1-dla-local} is
\begin{eqnarray}
Q_1(\eta,\rho) = Q_1^{(0)}\,I_0\!\left(2\sqrt{\bar\alpha_s\,\eta\,\rho}\right),
\qquad
\eta,\rho\ge 0.
\label{Q1-solution-I0}
\end{eqnarray}
Since $I_0(0)=1$, eq.~\eqref{Q1-solution-I0} reproduces the boundary conditions
\eqref{Born-bc}, while direct differentiation gives
\begin{eqnarray}
{\partial^2\over \partial\eta\,\partial\rho}Q_1(\eta,\rho) =
\bar\alpha_s\,Q_1(\eta,\rho).
\end{eqnarray}

For large $\eta\rho$, using
\begin{eqnarray}
I_0(x)\simeq \frac{e^x}{\sqrt{2\pi x}}
\qquad (x\to\infty),
\end{eqnarray}
one finds
\begin{eqnarray}
Q_1(\eta,\rho) \simeq Q_1^{(0)}
\frac{\exp\!\left(2\sqrt{\bar\alpha_s\,\eta\,\rho}\right)}
{\sqrt{4\pi\sqrt{\bar\alpha_s\,\eta\,\rho}}}.
\label{Q1-asymptotic}
\end{eqnarray}
Thus, before imposing any relation between $\rho$ and $\eta$, the resummation generated by eq.~\eqref{Q1-dla-local} is a mixed longitudinal--transverse double logarithm,
\begin{eqnarray}
(\bar\alpha_s\,\eta\,\rho)^n \sim\left(\alpha_s \ln{1\over x}\,\ln Q_\perp^2\right)^n,
\end{eqnarray}
rather than a pure double logarithm of energy.

\subsubsection{Longitudinally constrained transverse phase space: pure double logarithm of energy.}

To obtain instead the genuine double logarithm of energy, the transverse phase space must
not be treated as independent of the longitudinal ordering variable. Since in the present
formalism the evolution is written in the approximation in which the fields are ordered
according to the Sudakov component $k^+$, the logarithmic longitudinal phase space is
naturally written as
\begin{eqnarray}
\int \frac{dk^+}{k^+}\, .
\label{dkplusmeasure}
\end{eqnarray}
Equivalently, one may introduce the dimensionless ratio $k^+/q^+$, since the virtual
photon carries the large $q^+$ component in the chosen frame.

For one ladder rung, the relevant kinematic constraint is
\begin{eqnarray}
k_\perp^2 \lesssim \frac{k^+}{q^+}\,s = 2P^-k^+,
\label{kt-bound}
\end{eqnarray}
where in the last step we used $s\simeq 2q^+P^-$. In coordinate space this corresponds to
\begin{eqnarray}
{1\over 2P^-k^+}\ \lesssim\ r_\perp^2\ \lesssim\ \frac{1}{Q^2}.
\label{r-bound}
\end{eqnarray}
In this case, the logarithmic phase space is no longer a longitudinal logarithm times an
independent transverse logarithm, but rather
\begin{eqnarray}
\int_{x_B q^+}^{q^+} {dk^+\over k^+}
\int_{1/(2P^-k^+)}^{1/Q^2}\frac{dr_\perp^2}{r_\perp^2}
=
\int_{x_B q^+}^{q^+} \frac{dk^+}{k^+}\,
\ln{2P^-k^+\over Q^2}.
\label{one-rung-energy}
\end{eqnarray}
Using 	$x_B={Q^2\over s}$, and $s\simeq 2q^+P^-$,
we obtain
\begin{eqnarray}
\ln{2P^-k^+\over Q^2} = \ln{k^+\over x_B q^+}\,,
\label{log-kplus}
\end{eqnarray}
and therefore
\begin{eqnarray}
\int_{x_B q^+}^{q^+} {dk^+\over k^+}\,
\ln{k^+\over x_B q^+} = \half\,\ln^2{1\over x_B}\,.
\label{pure-energy-log}
\end{eqnarray}
This shows that, once the transverse bound is tied to the longitudinal variable $k^+$,
the second logarithm becomes again a logarithm of energy. In other words, the transverse
logarithm is not lost, but it is converted into a second longitudinal logarithm through the
kinematic constraint.

In this regime, $\rho$ is no longer an independent external variable. Rather, the maximal
transverse logarithm is fixed by the longitudinal phase space itself:
\begin{eqnarray}
\rho_{\rm max}(k^+)
\sim
\ln\frac{2P^-k^+}{Q^2} = \ln\frac{k^+}{x_B q^+}.
\label{rho-max}
\end{eqnarray}
Thus the appropriate pure-energy double-logarithmic regime corresponds to evaluating the
mixed solution \eqref{Q1-solution-I0} on a kinematic trajectory for which $\rho$ is of the
same order as $\eta$. In the symmetric double-logarithmic region, $\rho\simeq\eta$, one
obtains
\begin{eqnarray}
Q_1(\eta,\rho=\eta) = Q_1^{(0)}\,I_0\!\left(2\sqrt{\bar\alpha_s}\,\eta\right),
\label{Q1-pure-energy}
\end{eqnarray}
so that for large $\eta$
\begin{eqnarray}
Q_1(\eta,\rho=\eta)
\simeq
Q_1^{(0)}
\frac{\exp\!\left(2\sqrt{\bar\alpha_s}\,\eta\right)}
{\sqrt{4\pi\sqrt{\bar\alpha_s}\,\eta}}.
\label{Q1-pure-energy-asymptotic}
\end{eqnarray}
Accordingly, the fixed-coupling exponent is
\begin{eqnarray}
\Delta = 2\sqrt{\bar\alpha_s} = \sqrt{\frac{2\alpha_s C_F}{\pi}},
\label{KL-ladder}
\end{eqnarray}
which coincides with the fixed-coupling Kirschner--Lipatov ladder value in the symmetric
double-logarithmic regime.

The important point is that this pure-energy double logarithm is not obtained by modifying
the Mellin analysis itself, but by changing the phase-space boundary conditions. If one
keeps $L_\perp^2$ independent of $\eta$, one resums the mixed double logarithms
$(\alpha_s \ln(1/x_B)\ln Q_\perp^2)^n$. If instead the transverse phase space is constrained
by the longitudinal ordering according to eqs.~\eqref{kt-bound}--\eqref{r-bound}, then the
same ladder produces the genuine double logarithms of energy
$(\alpha_s \ln^2(1/x_B))^n$.

The solution \eqref{Q1-solution-I0} therefore makes explicit the relation between the
present high-energy evolution of $Q_1$ and the standard mixed double-logarithmic regime
in which one resums powers of $(\alpha_s \ln(1/x_B)\ln Q_\perp^2)^n$. In this sense, the
Bessel-type functional form obtained from eq.~\eqref{Q1-dla-local} is the natural mixed
longitudinal--transverse DLA solution associated with the unconstrained phase space.

By contrast, the genuine Kirschner--Lipatov double logarithms of energy,
$(\alpha_s \ln^2(1/x_B))^n$, are recovered only after imposing the longitudinally
constrained transverse phase space discussed above. In this regime, the second logarithm
is no longer generated by an independent external variable $\rho$, but by the kinematic
restriction which ties the transverse integration domain to the longitudinal ordering. The
fixed-coupling exponent \eqref{KL-ladder} is then obtained in the symmetric
double-logarithmic regime.

We emphasize, however, that in the present work we do not derive a separate closed
evolution equation directly in the single variable $\xi=\ln(1/x_B)$. Rather, we identify
the kinematic regime in which the longitudinally constrained transverse phase space
converts the second logarithm into a logarithm of energy, and in this symmetric
double-logarithmic regime we recover the corresponding ladder exponent.

Thus, the present formalism clarifies the relation between
the operator evolution of $Q_1$ and the Kirschner-Lipatov double-logarithmic
quark-exchange formalism~\cite{Kirschner:1982qf,Kirschner:1983di}. More precisely, with
independent transverse phase space one obtains the mixed longitudinal-transverse DLA
encoded in eq.~\eqref{Q1-dla-local}; after imposing the longitudinally dependent
transverse bound, one recovers the genuine double logarithm of energy and, in the
symmetric regime, the fixed-coupling exponent \eqref{KL-ladder}. This is the same ladder
exponent that appears in the non-singlet channel, while keeping the full finite-$N_c$ color
factor $C_F$ rather than only its strict large-$N_c$ limit $N_c/2$~\cite{Kovchegov:2018znm, Kovchegov:2017lsr, Kovchegov:2015pbl,Borden:2023ugd,Borden:2025ehe}.

In the polarized sector, the same reasoning applies to the helicity operator $Q^{{\rm NS},(a)}_5$, while for 
the singlet $Q^{\rm S}_5$ we need to consider the mixing with gluon operator. 
Thus, in the strict ladder approximation, the evolution of $Q^{{\rm NS},(a)}_5$ has the same mixed
double-logarithmic functional form as \eqref{Q1-solution-I0}, and after imposing the
longitudinally constrained transverse phase space it leads to the same square-root exponent
\eqref{KL-ladder}. Differences with the full Bartels-Ermolaev-Ryskin~\cite{Bartels:1995iu,Bartels:1996wc} result are expected
only beyond the strict ladder approximation, where signature-dependent and non-ladder
contributions become essential~\cite{Kovchegov:2016zex,Kovchegov:2016weo}. The comparison with the full BER result is left for future work.

\subsubsection{Evolution equation with kinematic constraint and its solution}

The discussion above shows that, once the transverse phase space is constrained by
longitudinal ordering, the natural variables are the effective rapidity $\eta$ and the
transverse logarithm $\rho$, with $\eta\ge 0$. It is then convenient to introduce also the
total longitudinal logarithm
\begin{eqnarray}
Y \equiv \eta+\rho\, ,
\qquad\qquad
Y\ge \rho\ge 0\, .
\label{Y-def}
\end{eqnarray}
The kinematic constraint does not define a new independent evolution equation. Rather, it
restricts the domain on which the mixed longitudinal-transverse DLA solution obtained
above has to be evaluated.

Starting from the integral form of the mixed DLA equation,
\begin{eqnarray}
Q_1(\eta,\rho) = Q_1^{(0)} + \bar\alpha_s\int_0^\eta d\eta_1\int_0^\rho d\rho_1\,Q_1(\eta_1,\rho_1),
\qquad
\eta\ge 0,\ \rho\ge 0,
\label{Q1-dla-integral-again}
\end{eqnarray}
we now rewrite it in terms of the variables
\begin{eqnarray}
Y_1 \equiv \eta_1+\rho_1 .
\label{Y1-def}
\end{eqnarray}
At fixed $\rho_1$ one has $d\eta_1=dY_1$, while the integration domain
\begin{eqnarray}
0\le \eta_1\le \eta,
\qquad
0\le \rho_1\le \rho,
\end{eqnarray}
becomes
\begin{eqnarray}
0\le \rho_1\le \rho,
\qquad
\rho_1\le Y_1\le Y-\rho+\rho_1.
\label{Y1-domain}
\end{eqnarray}
Indeed, the lower bound follows from $\eta_1\ge 0$, so $Y_1\ge \rho_1$
while the upper bound is simply the condition $\eta_1\le \eta$ so, $Y_1\le Y-\rho+\rho_1$.
Therefore, defining
\begin{eqnarray}
Q_1^{\rm kc}(Y,\rho)\equiv Q_1(Y-\rho,\rho),
\label{Q1kc-def}
\end{eqnarray}
we can write the mixed DLA equation in the nonlocal form
\begin{eqnarray}
Q_1^{\rm kc}(Y,\rho) = Q_1^{(0)} + \bar\alpha_s\int_0^\rho d\rho_1\int_{\rho_1}^{\,Y-\rho+\rho_1} dY_1\,
Q_1^{\rm kc}(Y_1,\rho_1),
\qquad
Y\ge \rho\ge 0.
\label{Q1kc-integral}
\end{eqnarray}

Differentiating eq.~\eqref{Q1kc-integral} with respect to $Y$ at fixed $\rho$, we obtain
\begin{eqnarray}
\frac{\partial}{\partial Y}Q_1^{\rm kc}(Y,\rho)
=
\bar\alpha_s
\int_0^\rho d\rho_1\,
Q_1^{\rm kc}(Y-\rho+\rho_1,\rho_1),
\qquad
Y\ge \rho\ge 0.
\label{Q1kc-nonlocal}
\end{eqnarray}
Since at fixed $\rho$ one has $\partial_Y=\partial_\eta$ and
$Q_1^{\rm kc}(Y,\rho)=Q_1(\eta,\rho)$ with $\eta=Y-\rho$, eq.~\eqref{Q1kc-nonlocal}
is nothing but the original local mixed DLA equation written in different variables:
\begin{eqnarray}
\frac{\partial}{\partial \eta}Q_1(\eta,\rho) =
\bar\alpha_s
\int_0^\rho d\rho_1\,
Q_1(\eta,\rho_1),
\qquad
\eta\ge 0,\ \rho\ge 0.
\label{Q1-local-again}
\end{eqnarray}
With the constant boundary conditions
\begin{eqnarray}
Q_1(0,\rho)=Q_1^{(0)},
\qquad
Q_1(\eta,0)=Q_1^{(0)},
\label{Q1-boundary-again}
\end{eqnarray}
the solution is therefore
\begin{eqnarray}
Q_1(\eta,\rho) =
Q_1^{(0)}\,I_0\!\left(2\sqrt{\bar\alpha_s\,\eta\,\rho}\right),
\label{Q1-solution-again}
\end{eqnarray}
or, equivalently, in terms of the constrained variables,
\begin{eqnarray}
Q_1^{\rm kc}(Y,\rho) = Q_1^{(0)}
\,I_0\!\left(2\sqrt{\bar\alpha_s\,(Y-\rho)\rho}\right),
\qquad
Y\ge \rho\ge 0.
\label{Q1kc-solution}
\end{eqnarray}

In the symmetric double-logarithmic regime,
\begin{eqnarray}
\rho=\eta
\qquad\Longleftrightarrow\qquad
\rho=\frac{Y}{2},
\label{symm-regime-Y}
\end{eqnarray}
one finds
\begin{eqnarray}
Q_1^{\rm kc}\!\left(Y,\frac{Y}{2}\right) =
Q_1^{(0)}\,I_0\!\left(\sqrt{\bar\alpha_s}\,Y\right)
\simeq
Q_1^{(0)}\,
\frac{\exp\!\left(\sqrt{\bar\alpha_s}\,Y\right)}
{\sqrt{2\pi\sqrt{\bar\alpha_s}\,Y}},
\label{Q1kc-symmetric}
\end{eqnarray}
which is exactly the same result as
\begin{eqnarray}
Q_1(\eta,\rho=\eta) = Q_1^{(0)}\,I_0\!\left(2\sqrt{\bar\alpha_s}\,\eta\right)
\label{Q1-symmetric-eta}
\end{eqnarray}
after identifying $Y=2\eta$.

Thus, the form \eqref{Q1kc-integral} does not define a new one-variable evolution problem.
It is simply the mixed DLA equation rewritten in variables adapted to the kinematically
restricted phase space. The genuine double logarithm of energy is recovered not by changing
the kernel, but by evaluating the same mixed DLA solution in the regime $\rho\sim \eta$
selected by the kinematic constraint. In particular, in the symmetric case \eqref{symm-regime-Y}
one recovers the fixed-coupling Kirschner--Lipatov exponent. The same reasoning applies to
$Q^{{\rm NS}, (a)}_5$ in the non-singlet ladder approximation.

\section{Conclusions}

In this work we studied how the familiar quark and helicity light-ray operators of DIS emerge from the 
high-energy expansion beyond the strict eikonal approximation. We 
showed that the first sub-eikonal quark correction to the dipole description already reconstructs,
in the inclusive limit, the standard quark and helicity operator content at 
nonzero Bjorken $x_B$. At the differential level, the same correction is governed by a quark TMD-like 
light-ray operator. In this sense, the first correction beyond the 
eikonal dipole approximation already contains the longitudinal operator structure needed for the partonic interpretation of DIS at finite $x_B$.

A key point of the analysis is that the small-$x_B$ approximation and the full phase-space integration 
do not commute. If the small-$x_B$ limit is taken too early, one 
recovers only the naive collinear operator at $x_B=0$. By contrast, when the phase-space integration is completed first,
the transverse kernel eikonalizes in the high-energy 
limit while the longitudinal Fourier phase remains exact, and the inclusive result reproduces the standard
nonlocal quark and helicity distributions at nonzero $x_B$. This 
makes explicit how the finite-$x_B$ light-ray structure emerges from the high-energy formalism already at first sub-eikonal order.

We further showed that the same inclusive operator content follows independently from the high-energy limit 
of the leading-twist non-local OPE of Balitsky and Braun. This 
provides a nontrivial cross-check of the construction and establishes an explicit operator-level bridge between 
the shock-wave formalism and the non-local light-cone 
expansion. The present analysis does not yet imply a general one-to-one correspondence 
between higher-twist order and sub-eikonal order. Rather, it provides a concrete 
framework in which that broader relation can be investigated systematically.

A second main result of this work is the discussion of the high-energy evolution of the corresponding 
$x_B=0$ operators $Q^{\rm S}_1$, $Q_1^{{\rm NS}, (a)}$,
$Q^{\rm S}_5$, and $Q_5^{{\rm NS}, (a)}$, and their Hermitian conjugates. 
This singlet/non-singlet decomposition is meant here at the level of the flavor structure of the $x_B=0$ high-energy operators and of 
their rapidity evolution, and should not be identified directly with the decomposition of the physical one-photon helicity structure function at finite $x_B$, whose 
contribution is weighted by the electromagnetic charges $e_f^2$.
Rewriting the evolution 
equations in terms of dipole-type operator combinations, eqs. \eqref{Q1xy}, \eqref{Q5xy}, \eqref{Psi1xy}, \eqref{Psi5xy}, \eqref{calFxy}, and \eqref{dipoleWilsonline},
 we identified an operator basis whose bilocal building blocks vanish in the zero-dipole-size limit. This makes the 
small-dipole behavior manifest, clarifies the operator structure responsible for the leading logarithm of energy, and provides a natural framework for organizing the 
sub-eikonal high-energy dynamics. 

Projecting the evolution equations onto the strict local ladder sector, one finds that
$Q_1^{NS,(a)}$, and $Q_5^{NS,(a)}$ admit the usual mixed
longitudinal-transverse Bessel-type solution when the transverse phase space is treated
independently. When the transverse phase space is instead constrained by longitudinal
ordering, the second logarithm is converted into a logarithm of energy. In the symmetric
double-logarithmic regime, this yields the fixed-coupling Kirschner-Lipatov exponent with
the full finite-$N_c$ color factor $C_F$. This statement refers to the reduced strict-ladder
sector, while the full enlarged operator basis is not closed in the present formulation.

It is worth stressing that the logarithmic high-energy enhancement discussed here was also observed 
in the collinear framework where the corresponding small-$x$ enhancement
was analyzed directly in higher-order DIS coefficient functions and splitting kernels~\cite{Vermaseren:2005qc,Davies:2022ofz}.

Taken together, the results in this work show that the first sub-eikonal quark correction already provides the operator bridge between the high-energy dipole formalism and 
the standard finite-$x_B$ light-ray description of DIS, while at the same time leading naturally to an $x_B=0$ evolution problem formulated in terms of dipole-type 
operators. A natural next step is therefore to derive a closed evolution system directly for this operator basis and to clarify in detail its relation to previous approaches to 
small-$x$ helicity evolution beyond the strict ladder approximation.

\section{Acknowledgments}

The author gratefully acknowledges financial support from the Theoretical Physics Division of the National Centre for Nuclear Research (NCBJ). Part of this work was 
developed during the workshop \emph{Bridging TMD Frameworks: Intersections, Tensions, and Applications} at ECT* in Trento. The author thanks ECT* for its hospitality and 
stimulating scientific atmosphere, and also acknowledges the support of ECT* and INFN during the workshop. The author thanks the University of Salento for its hospitality 
during the completion of this work. The author is grateful to I.~Balitsky, Yu.~Kovchegov, and A.~Vladimirov for useful discussions.
	
\appendix

\section{Diagram in Fig. \ref{Fig:DIS-q-1pointSW}b}
\label{sec:FigDIS-q-1pointSWb}

Let us calculate diagram in Fig. \ref{Fig:DIS-q-1pointSW}b. Our starting point is
\begin{eqnarray}
	&&\langle \bar{q}(k)|\gamma^*(q)\rangle
	\nonumber\\
	=\!\!\!&& iee_f\Big(\int d^4x e^{-iq\cdot x}\varepsilon_\mu(q)\Big)\int d^4y\,\langle {\rm T}\{\barpsi(y)\barpsi(x)\gamma^\mu\psi(x)\}\rangle
	(-i\overleftarrow{\slashd}_y)v(k)\theta(k^+)e^{ik\cdot y}
	\nonumber\\
	\hspace{-2cm}
	&&\times\!\int d^4z\, e^{-iq\cdot z}\varepsilon_\nu(q)[-i\tildeD(q)]^\nu_{~\alpha}\langle{\rm T}\{A^\alpha(z)A_\mu(x)\}\rangle
\end{eqnarray}
the differentiation brings again two terms
\begin{eqnarray}
	&&\langle \bar{q}(k)|\gamma^*(q)\rangle
	\nonumber\\
	=\!\!\!&& - ee_f\left(\int d^4x \,e^{-iq\cdot x}\varepsilon_\mu(q)\right)\int d^4y\,\barpsi(x^+,x_\perp)\gamma^\mu
	\int{\dhd^4 k_2\,e^{-ik_2\cdot(y-x)}\over 2k_2^+(k^2_2+i\epsilon)}\nonumber\\
	&&	\times\Bigg[
	\left([x^+,-\infty n_1]_x\ssk_2 - g\int_{-\infty}^{x^+}dw^+\gamma^i[x^+,w^+]_x {F_i}^{\;-}[w^+,-\infty n_1]_x\right)
	\nonumber\\
	&&\times
	\ssn_2\ssk_2\Big(i\,\ssn_2\delta(x^+-y^+) + \ssk_2\theta(x^+ - y^+)\Big)
	\nonumber\\
	&&+ \left([x^+, +\infty n_1]_x\ssk_2 + g\int^{\infty}_{x^+}dw^+\gamma^i[x^+,w^+]_x {F_i}^{\;-}[w^+,+\infty n_1]_x\right)\ssn_2\ssk_2
	\nonumber\\
	&&\times\Big(-i\,\ssn_2\delta(x^+-y^+)+\ssk_2\theta(y^+-x^+)\Big)
	\Bigg]v(k)\,e^{ik\cdot y}\theta(k^+)
	\nonumber\\
\end{eqnarray}
Integrating over $d^4y$ and $\dhd k^+\dhd^2k$, we arrive at
\begin{eqnarray}
	&&\langle \bar{q}(k)|\gamma^*(q)\rangle\theta(k^+)
	\nonumber\\
	=\!\!\!&& - {ee_f\over 2}\left(\int d^4x \,e^{-iq\cdot x}\varepsilon_\mu(q)\right)\int\dhd k_2^-
	{e^{ik^+ x^- + ik^- x^+  -i(k,x)}\over k^+ (2k^+k_2^-+k^2_\perp-i\epsilon)}\barpsi(x^+,x_\perp)\gamma^\mu 
	\nonumber\\
	&&\times\Bigg[\Big([x^+,-\infty n_1]_x (k^+\ssn_1+\ssk_\perp) 
	+ g\int_{-\infty}^{x^+}dw^+\gamma^i[x^+,w^+]_x {F_i}^{\;-}[w^+,-\infty n_1]_x\Big)
	\nonumber\\
	&&\times\, i\,\ssn_2(k^+\ssn_1+\ssk_\perp)\Big(\ssn_2 + {k^+\ssn_1- k_2^-\ssn_2 +\ssk_\perp\over k_2^-+k^- - i\epsilon}\Big)
	\nonumber\\
	&& + \Big([x^+,+\infty n_1]_x (k^+\ssn_1+\ssk_\perp) - g\int^{\infty}_{x^+}dw^+\gamma^i[x^+,w^+]_x {F_i}^{\;-}[w^+, +\infty n_1]_x\Big)
	\nonumber\\
	&&\times\,(-i)\ssn_2(k^+\ssn_1+\ssk_\perp)\Big(\ssn_2 + {k^+\ssn_1-k_2^-\ssn_2 +\ssk_\perp\over k_2^- + k^- + i\epsilon}\Big)
	\Bigg]v(k)
\end{eqnarray}
Finally, we take the residue integrating over $k^-_2$ and arrive at
\begin{eqnarray}
	&&\hspace{-1.4cm}\langle \bar{q}(k)|\gamma^*(q)\rangle
	\nonumber\\ 
	&&\hspace{-1.8cm}= - \dbar(q^+ - k^+)ee_f\,{\theta(k^+)\over 2k^+}
	\int \!\! d^2xdx^+ e^{-i(q^- - k^-)x^+ + i(q-k,x)_\perp}
	\barpsi(x^+,x_\perp)\sslash{\varepsilon}(q)
	\nonumber\\
	&&\hspace{-1.4cm}\times\Big([x^+, +\infty n_1]_x(k^+ \ssn_1+\ssk_\perp) - g\!\int_{x^+}^{+\infty} \!dw^+\gamma^i[x^+,w^+]_x {F_i}^{\;-}
	[w^+,+\infty n_1]_x\Big)\ssn_2v(k)
\end{eqnarray}

\section{From gauge-link to light-cone gauge link}
\label{sec:straightgaugelink}

Consider the effect of a large longitudinal boost parameter $\lambda$ on the components of the gauge fields. We have
\begin{eqnarray}
	&&A^-(x^-, x^+, x_\perp) \to \lambda\, A^-(\lambda^{-1}x^-, \lambda\, x^+, x_\perp)\,,\nonumber\\
	&&A^+(x^-, x^+, x_\perp) \to  \lambda^{-1}A^+(\lambda^{-1}x^-, \lambda\, x^+, x_\perp)\,,
	\label{boost}\\
	&&A_\perp(x^-, x^+, x_\perp)  \to  A_\perp(\lambda^{-1}x^-, \lambda\, x^+, x_\perp)\,,\nonumber
\end{eqnarray}
and the field strength is rescaled as follows
\begin{eqnarray}
&&{F_i}^{\;-}(x^-, x^+, x_\perp) \to  \lambda\, {F_i}^{\;-}(\lambda^{-1}x^-, \lambda\, x^+, x_\perp)\,,\nonumber\\
&&{F_i}^{\;+}(x^-, x^+, x_\perp) \to  \lambda^{-1}{F_i}^{\;+}(\lambda^{-1}x^-, \lambda\, x^+, x_\perp)\,,
\label{Fboost}\nonumber
\\
&&F^{-+}(x^-, x^+, x_\perp)  \to  F^{-+}(\lambda^{-1}x^-, \lambda\, x^+, x_\perp)\,,
\nonumber\\
&&F_{ij}(x^-, x^+, x_\perp)  \to  F_{ij}(\lambda^{-1}x^-, \lambda\, x^+, x_\perp)\,.
\end{eqnarray}
For the spinor fields we have
\begin{eqnarray}
\bar{\psi}t^a\ssn_1\psi \to  \lambda \bar{\psi}t^a\ssn_1\psi\,,~~~~
\bar{\psi}t^a\gamma^\perp_\nu\psi \to  \bar{\psi}t^a\gamma^\perp_\nu\psi\,,~~~~
\bar{\psi}t^a\ssn_2\psi \to  \lambda^{-1} \bar{\psi}t^a\ssn_2\psi\,.
\label{spinorboost}
\end{eqnarray}

Let us consider the effect of the large longitudinal boost on a straight gauge link connecting the two points $x^\mu$ and $y^\mu$, defined as
\begin{eqnarray}
[x,y] = \mathrm P\exp\!\left\{ig\int_y^x dz^\mu\,A_\mu(z)\right\}\,,
\label{gaugelink}
\end{eqnarray}
where we parameterize the straight path as
\begin{eqnarray}
z^\mu(u)=ux^\mu+\bar u\,y^\mu,\qquad u\in[0,1],\qquad \bar u\equiv 1-u.
\label{straightpath}
\end{eqnarray}
Decomposing in light-cone variables,
\begin{eqnarray}
v^\mu=v^+n_1^\mu+v^-n_2^\mu+v_\perp^\mu,
\end{eqnarray}
we have along the path
\begin{eqnarray}
&&z^+(u)=ux^+ + \bar u\, y^+,\qquad z^-(u)=ux^- + \bar u\, y^-\,,
\nonumber\\
&&z_\perp(u)=ux_\perp+\bar u\,y_\perp.
\label{zperp-linear}
\end{eqnarray}
Using $dz^\mu=(x-y)^\mu du$, the gauge link can be written as
\begin{eqnarray}
[x,y]=\!\!\!&& \mathrm P\exp\Bigg\{ig\int_0^1 du\,\Big[(x^+-y^+)A^-(z(u))
\nonumber\\
&&\hspace{1.3cm}
+(x^--y^-)A^+(z(u)) + (x_\perp-y_\perp)^iA_i(z(u))\Big]\Bigg\},
\label{gaugelink-lc}
\end{eqnarray}
After a large longitudinal boost of the target, the dominant gauge-field component is $A^-$,
while the contributions of $A^+$ and $A_i$ are power suppressed. Keeping only the leading term in the high-energy expansion, the straight gauge link reduces to
\begin{eqnarray}
[x,y]=\!\!\!&& \mathrm P\exp\left\{ig\int_{y^+}^{x^+}dz^+\,A^-(z^+,z^-(z^+),z_\perp(z^+))\right\}+ O(\lambda^{-1})
\label{gaugelink-boost1}
\end{eqnarray}
We can make further expansion and the dependence on $z^-$ becomes irrelevant at leading power, so that
\begin{eqnarray}
[x,y]=\!\!\!&& \mathrm P\exp\left\{ig\int_{y^+}^{x^+}dz^+\,A^-(z^+,z_\perp(z^+))\right\} + O(\lambda^{-1})\,.
\label{gaugelink-boost2}
\end{eqnarray}
Thus, at leading power, the straight gauge link is replaced by a Wilson line ordered along
$z^+$, built from the dominant field component $A^-$. At this point, however, the
Wilson line still follows the straight transverse interpolation
$z_\perp(z^+)=u(z^+)x_\perp+\bar u(z^+)y_\perp$.

If we now expand in the transverse separation $r_\perp=x_\perp-y_\perp$, we may
replace the transverse argument by a fixed coordinate, for instance $x_\perp$, up to subleading
corrections:
\begin{eqnarray}
\hspace{-0.7cm}A^-(z^+,z_\perp(z^+))=\!\!\!&& A^-(z^+,x_\perp)
+\big(z_\perp(z^+)-x_\perp\big)^i \partial_i A^-(z^+,x_\perp)
+O\!\left((z_\perp-x_\perp)^2\right).
\label{Aminus-expand}
\end{eqnarray}
In this way, the straight Wilson line reduces to the usual eikonal Wilson line at fixed
transverse position, while the omitted terms are subleading and generate the first
corrections associated with transverse gradients and, after gauge-covariant rearrangement,
field-strength insertions.

\bibliographystyle{apsrev4-2}
\bibliography{MyReferences}

\end{document}